\documentclass[a4paper,11pt]{article}
\usepackage{pos}
\usepackage{url}

\usepackage{setspace}

\title{
Properties, ensembles and hadron spectra with Stabilised Wilson Fermions
}
\ShortTitle{Properties, ensembles and hadron spectra with Stabilised Wilson Fermions}

\dedicated{
\vspace{3ex}\\
\includegraphics[width = 0.22\textwidth]{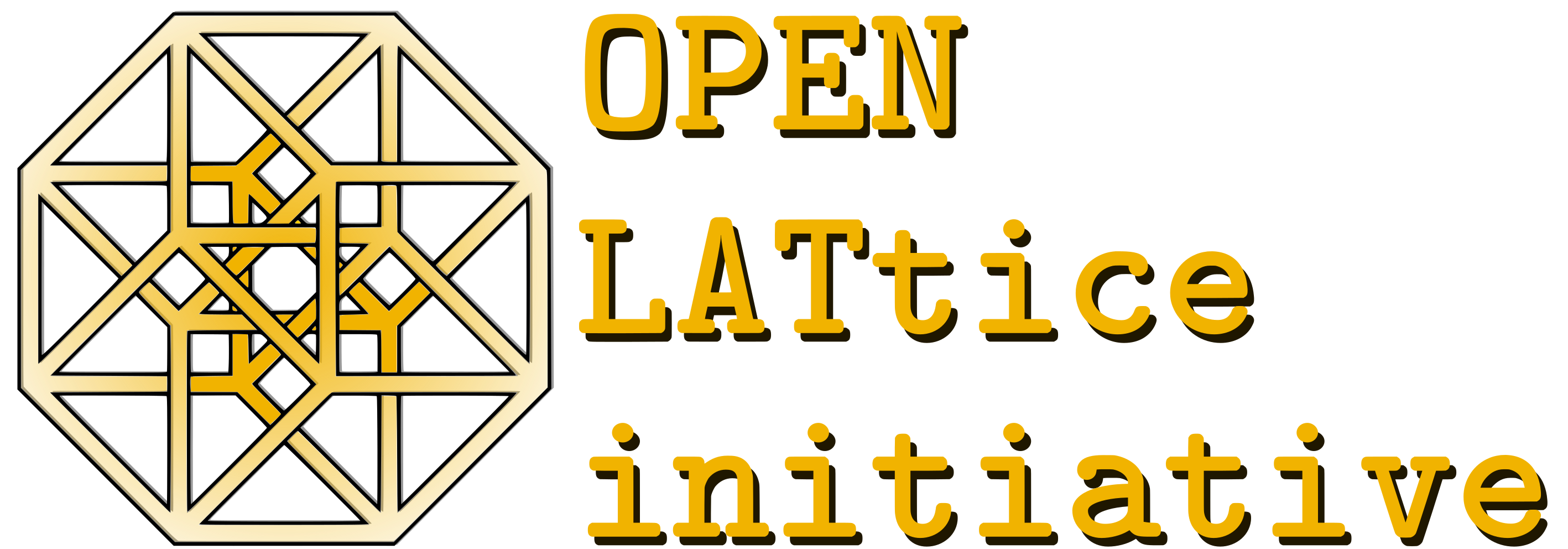}
\vspace{3ex}
}

\author[a]{F. Cuteri}
\author*[b,c,d,1]{A. Francis}
\author[e]{P. Fritzsch}
\author*[f,2]{G. Pederiva}
\author[g,d]{A. Rago}
\author[f]{A. Shindler}
\author[h]{A. Walker-Loud}
\author[i]{S. Zafeiropoulos}

\affiliation[a]{Institut für Theoretische Physik, Goethe Universität
,  Max-von-Laue-Str. 1, 60438 Frankfurt, Germany}
\affiliation[b]{Albert Einstein Center, Universität Bern, Sidlerstrasse 5, 3012 Bern, Switzerland}
\affiliation[c]{Institute of Physics, National Yang Ming Chiao Tung University, 30010 Hsinchu, Taiwan}
\affiliation[d]{Theory Department, CERN,
Esplanade des Particules 1, 1201 Geneva, Switzerland}
\affiliation[e]{School of Mathematics, Trinity College Dublin, Dublin 2, Ireland}
\affiliation[f]{FRIB \& Physics Department,
Michigan State University, East Lansing, MI 48824, USA}
\affiliation[g]{Centre for Mathematical Sciences, Plymouth University, Plymouth, PL4 8AA, United Kingdom}
\affiliation[h]{Nuclear Science Division, Lawrence Berkeley National Laboratory, Berkeley, CA 94720, USA}
\affiliation[i]{Aix Marseille Univ, Universit\'e de Toulon, CNRS, CPT, Marseille, France.}

\abstract{

 In this joint contribution we announce the formation of the \textit{OPEN LATtice initiative}, https://openlat1.gitlab.io, to study Stabilised Wilson Fermions (SWF). They are a new avenue for QCD calculations with Wilson-type fermions and we report results on our continued study of this framework: Tuning the clover improvement coefficient, and extending the reach of lattice spacings to $a=0.12$~fm. We fix the flavor symmetric points $m_\pi=m_K=412$~MeV at $a=0.055,0.064, 0.077, 0.094, 0.12$~fm and define the trajectories to the physical point by fixing the trace of the quark mass matrix. Currently our pion mass range extends down to $m_\pi\sim200$~MeV. We outline our tuning goals and strategy as well as our future planned ensembles. First scaling studies are performed on $f_\pi$ and $m_\pi$. Additionally results of a preliminary continuum extrapolation of $m_N$ at the flavor symmetric point are presented. 
Going further a first determination of the light and strange hadron spectrum chiral dependence is shown, which serves to check the quality of the action for precision measurements. We also investigate other quantities such as flowed gauge observables to study how the continuum limit is approached.
Taken together we observe the SWF enable us to perform stable lattice simulations across a large range of parameters in mass, volume and lattice spacing.\\
Pooling resources our new initiative has made our reported progress possible and through it we will share generated gauge ensembles under an open science philosophy.

}

\FullConference{%
 The 38th International Symposium on Lattice Field Theory, LATTICE2021
  26th-30th July, 2021
  Zoom/Gather@Massachusetts Institute of Technology
}

\begin{document}

\maketitle

\section{Introduction}

Wilson-Clover fermions (WCF) \cite{Sheikholeslami:1985ij}
have been and continue to be one of the most popular fermion discretisations in use in the lattice gauge theory community. Among their many attractive features they are conceptually clear to work with, there are many advanced methods and public codes available, they are relatively cheap and pose little restrictions on which observables can be computed. 
Together with a rigorous improvement program they can also be $\mathcal{O}(a)$-improved, where $a$ denotes the lattice spacing, thereby removing one of their most visible drawbacks. 
This makes them a flexible tool for carrying out a variety of physics programs, see e.g. the most recent FLAG report \cite{Aoki:2021kgd} for a non-exhaustive collection of fields where WCF have impacted. 
Nevertheless, there are some drawbacks that limit studies with WCF, as e.g., without automatic $\mathcal{O}(a)$ improvement\footnote{An alternative, well-established discretisation is Wilson twisted mass (WTM) at maximal twist~\cite{Frezzotti:2000nk, Frezzotti:2005gi}. The drawback for WTM is the breaking of parity and flavor symmetries at finite lattice spacing and the constraint of an even number of quark flavors. For a review see Ref.~\cite{Shindler:2007vp}.} observables often require finer $a$ than other actions as higher order effects can become a difficulty. Generating gauge fields at fine $a$ in turn requires dealing with topology freezing problems and critical slowing down. 
Furthermore without chiral symmetry the lowest eigenvalue of the Dirac operator is not protected from taking arbitrarily small values. This can become a problem especially when the lattice spacing is coarse or the pion mass light. 
%
These features limit the parameter space where WCF simulations can be successfully and safely deployed. They define the window of possibility in lattice spacing, volume and quark mass in which all current simulations take place\footnote{Such windows of possibility can be defined for all lattice actions. Many bounds are indeed similar across all.}. 
%

Taming the large volume limitation of WCF to enable master-field simulations \cite{Luscher:2017cjh} lead to a reformulation of WCF~\cite{Francis:2019muy} that incorporates several numerical stabilising techniques but also a local change of the fermion action - the original clover term $c_{SW}$ being replaced with an exponentiated version of it. The combined package of algorithmic and fermionic stabilising measures are called stabilised Wilson fermions (SWF).

First results from deploying the SWF framework in the master-field context have been reported at this meeting \cite{Ce:2021akh,Fritzsch:2021klm} demonstrating the effectiveness of SWF in removing the large volume limitation.
Aside of the volume aspects SWF incorporate further benefits applicable to all types of simulation. For example, smaller values of $c_{SW}$ at a given lattice spacing with SWF were seen compared to WCF. Overall good continuum scaling and relative $am_q$ effects in the limited studies performed were also observed. 
Even though it is too early to draw firm conclusions, the opportunity to work with smaller masses at coarser lattice spacings makes it interesting to continue studying SWF in the traditional volume setup.
If confirmed, these features could be particularly attractive for nuclear and nucleon applications, as well as $(g-2)_\mu$ and many of those calculations entering the FLAG report. 
For this reason we founded the \textit{OPEN LATtice initiative}, \href{https://openlat1.gitlab.io}{https://openlat1.gitlab.io}, which has made the following new studies possible and whose goals we detail in the last part of this proceedings contribution.
%

\section{Stabilised Wilson fermion toolkit}

Simulations with SWF imply the use of several measures designed to lead to a more stable generation of gauge fields. As such SWF are a collection of tools, as opposed to one single change, and some of their components have been established previously (e.g. the stochastic molecular dynamics (SMD) shown below). 
Roughly, the measures combined in the SWF can be split into two categories: Those aiming at an improved algorithmic stability, and those aiming at improving aspects of the fermion discretisation. 
We note that the tools presented here are used in addition to established techniques, such as SAP, local deflation, multi-grid, mass-preconditioning, multiple time-scale integrators etc., in our simulations.
All of these methods are implemented within the open source software package \verb+openQCD-2.0+ \cite{mluscher:openqcd}. 

\subsection{Improving algorithmic stability}

The first ingredient is to increase the stability in MD evolution in the gauge field generation process. One aspect to note here is that in the HMC approach large jumps in the phase space trajectory can occur, due to accumulated integration errors for example. Once they happen some re-thermalisation is required to return to sampling the target distribution, which can lead to extended autocorrelation times.
An alternative approach is to switch to the related SMD algorithm \cite{Horowitz:1985kd,Horowitz:1986dt,HOROWITZ1991247,Jansen:1995gz}. 
In the SMD an update cycle is schematically given by: \\
\begin{center}
\begin{tabular}{cll}
\hline
\hline
1. & Refresh $\pi(x,\mu)$ and $\phi(x)$ by a random field rotation:& $\pi \rightarrow c_1\pi + c_2 v$\\
& &$\phi \rightarrow c_1\phi + c_2 D^\dagger\eta$\\
& & ($v$ and $\eta$ normal distributed)\\
&$c_1^2 + c_2^2 =1$, $c_1 =e^{-\epsilon \gamma}$,~~~~ 
$\epsilon=$ MD integration time,&$\gamma=$ friction parameter\\
2. & short MD evolution &\\
3. & Accept/Reject-step & (exact algorithm)\\
4. & Repeat $\circlearrowleft$ &\\
\hline
\hline
\end{tabular}
\end{center}
Here $\pi(x,\mu)$, $\phi(x)$ and $U(x,\mu)$ denote the momentum and pseudofermion fields as well as gauge links, respectively.
The SMD is an exact algorithm that coincides with the HMC at fixed $\epsilon$ and large $\gamma$. For small $\epsilon$ the SMD can be shown to be ergodic and to converge
to a unique stationary state simulating the canonical distribution \cite{mluscher:note}.
The SMD gives an effective reduction of unbounded energy violations $|\delta H|\gg 1$ and exhibits shorter autocorrelation times
\cite{Luscher:2011kk,Luscher:2017cjh}, largely compensating for the longer time per MDU required compared to the HMC.
Note that we use the version of the SMD with the accept/reject step included.
When configurations are rejected the momentum is reversed and the trajectory tends to backtrack with a period $t_{acc}=\delta \tau P_{acc}/(1-P_{acc})$ \cite{Luscher:2011kk}. As a result rejections should ideally occur only at large distances in the evolution, mandating a high acceptance rate typically in excess of $98\%$.
In passing we remark that the smooth changes in $\phi_t$ and $U_t$ improve the update of the deflation subspace.
Finally, note that since $\delta H\propto\sqrt{V}$, higher integration rules should be used to increase integration precision as the volume is increased. 
The second ingredient for an improved stability is to utilize a volume-independent norm for the solver stopping criterion:
\begin{center}
\begin{tabular}{ll}
\hline
\hline
$\Vert \eta - D \tilde\psi \Vert_2 \leq w \Vert \eta \Vert_2$, & $\Vert \eta \Vert_2 = \Big(\sum_x (\eta(x),\eta(x)) \Big)^{1/2}\propto \sqrt{V}$\\
uniform norm: & $\Vert \eta \Vert_\infty=\textrm{sup}_x \Vert \eta \Vert_2$, V-independent\\
\hline
\hline
\end{tabular}
\end{center}
this norm guarantees the quality of a given solve and gives insurance against precision losses from local effects.
As final algorithmic ingredient, note that for the global accept/reject step $\delta H\propto \epsilon^P\sqrt{V}$. This can lead to accumulation errors for global sums and to remedy this issue quadruple precision is implemented in \verb+openQCD-2.0+.

\subsection{The exponentiated Clover action}

Next we turn to measures aimed at improving aspects of the fermion discretisation. This marks a departure from the standard WCF setup and defines a new action. 
To start, recall the $\mathcal{O}(a)$-improved Wilson Dirac operator:
\begin{equation}
D = \frac{1}{2} \Big[\, \gamma_\mu\Big( \nabla_\mu^*+\nabla_\mu - a\nabla_\mu^*\nabla_\mu \Big)\,\Big] + c_{SW} \frac{i}{4}\sigma_{\mu\nu}\hat F_{\mu\nu} + m_0~~.
\end{equation}
Typically one next classifies the lattice points as even/odd and writes
the preconditioned form,
$\hat D = D_{ee} - D_{eo} ( D_{oo} )^{-1} D_{oe}$
with diagonal part ($M_0=4 +m_0$):
\begin{equation} 
D_{ee} + D_{oo} = M_0 + c_{SW} \frac{i}{4} \sigma_{\mu\nu} \hat F_{\mu\nu} ~~. \end{equation}
In this form the Dirac operator is not protected from arbitrarily small eigenvalues originating from the second, clover, term. In particular, the clover term can saturate the bound
$\Vert \frac{i}{4} \sigma_{\mu\nu} \hat F_{\mu\nu} \Vert_2 \leq 3$
while the clover coefficient $c_{SW}$ at tree-level is one and then grows monotonically with $g_0^2$.
Furthermore, the positive and negative eigenvalues of the clover term are equally distributed.
Taken together this makes the above statement more precise as we see that $D_{oo}$ is not protected from arbitrarily small eigenvalues. This effect becomes more pronounced in simulations with small quark masses or rough, coarse gauge fields or large lattices. As the probability to encounter such an arbitrarily small eigenvalue of $D_{oo}$ increases with the volume, large volume simulations become pathological.
%
To remedy this situation the suggestion is to use a different form of the clover term, one that is bounded from below by construction (for further details see \cite{Francis:2019muy}):
\begin{equation} 
D_{ee} + D_{oo} = M_0 + c_{SW} \frac{i}{4} \sigma_{\mu\nu} \hat F_{\mu\nu} ~\rightarrow~~ M_0\exp\Big[ \frac{c_{SW}}{M_0}\frac{i}{4} \sigma_{\mu\nu} \hat F_{\mu\nu}  \Big]~~. \end{equation}
One can convince oneself that this form is valid in terms of Symanzik improvement. It guarantees invertibility of the clover term, one of the features that is not present in the standard setup. 

\subsection{The stabilised Wilson fermion package}

Once more, we understand stabilised Wilson fermions as the combination of all the above mentioned measures into one simulation setup. 
The algorithmic measures are not unique to the exponentiated Clover action and can be used with other actions too. 
Some of the measures presented here show increasing benefit as the volume of a simulation increases.  

Note that in the SWF setup the suggestion is to introduce a local change to the action only\footnote{In particular, the reformulation of the WCF with the exponentiated Clover does not amount to a type of smearing.}. 
The exponentiated clover action preserves the perturbative, Symanzik, expansion, which is particularly important for renormalisation. Employing the exponentiated clover could be interesting for other Wilson-type fermion discretisations as well for this reason. 

While SWF increase the accessible parameter window for lattice simulations, as we will see below, they do not cancel the drawbacks of Wilson fermions in general. For example, the Dirac operator can still exhibit exceptionally small eigenvalues, although their source will not be the clover term anymore. The SWF are not a cure for this problem but alleviate it for certain parameter regions. Finding out where the edges of the parameter window are is part of our continuing motivation to study SWF.

\section{Exploratory studies in quenched QCD}

Before passing to full QCD we study the effects of the exponentiated clover action in quenched QCD, i.e. $SU(3)$ gauge fields with valence quark probes. 
As in full QCD the improvement coefficients $c_{SW}$ and $c_A$ need to be tuned non-perturbatively.
The results of the tuning and the quenched hadron spectra provide some valuable insights into the exponential clover action with some caveats: Due to the special properties of the quenched theory the behaviour of the lowest Dirac eigenvalues cannot be understood as rigorously as in the full QCD case, for example. Nevertheless, we present results here as testbed for the exponentiated clover idea going towards full QCD.

In addition, even though quenched simulations for precision QCD observables are phasing out, they are still commonly used\footnote{The references here are just a few non-exhaustive examples chosen for their recent emergence.} for exploratory studies of new observables \cite{Orginos:2017kos,Shanahan:2020zxr}, proof-of-concept developments of new methods \cite{Shindler:2015aqa,Altenkort:2020fgs} or in the context of BSM models \cite{Lucini:2021xke} and high temperature QCD \cite{Brambilla:2020siz,Altenkort:2020axj}.
The presented studies, providing non-perturbatively tuned improvement coefficients and insights into spectral behaviour, open up the benefits of SWF to these efforts as well.

\subsection{Non-perturbative tuning of $c_{SW}$ and $c_A$}

To tune the improvement coefficients $c_{SW}$ and $c_A$ we follow the well established procedure in the Schr\"odinger functional scheme \cite{Luscher:1996ug} (SF) and take over notation from there. Throughout this section the lattice size is $L=8$ and $T=16$ with $N=2000$ independent configurations.  

\begin{figure}[t!]
\centering
\includegraphics[width=0.32\textwidth]{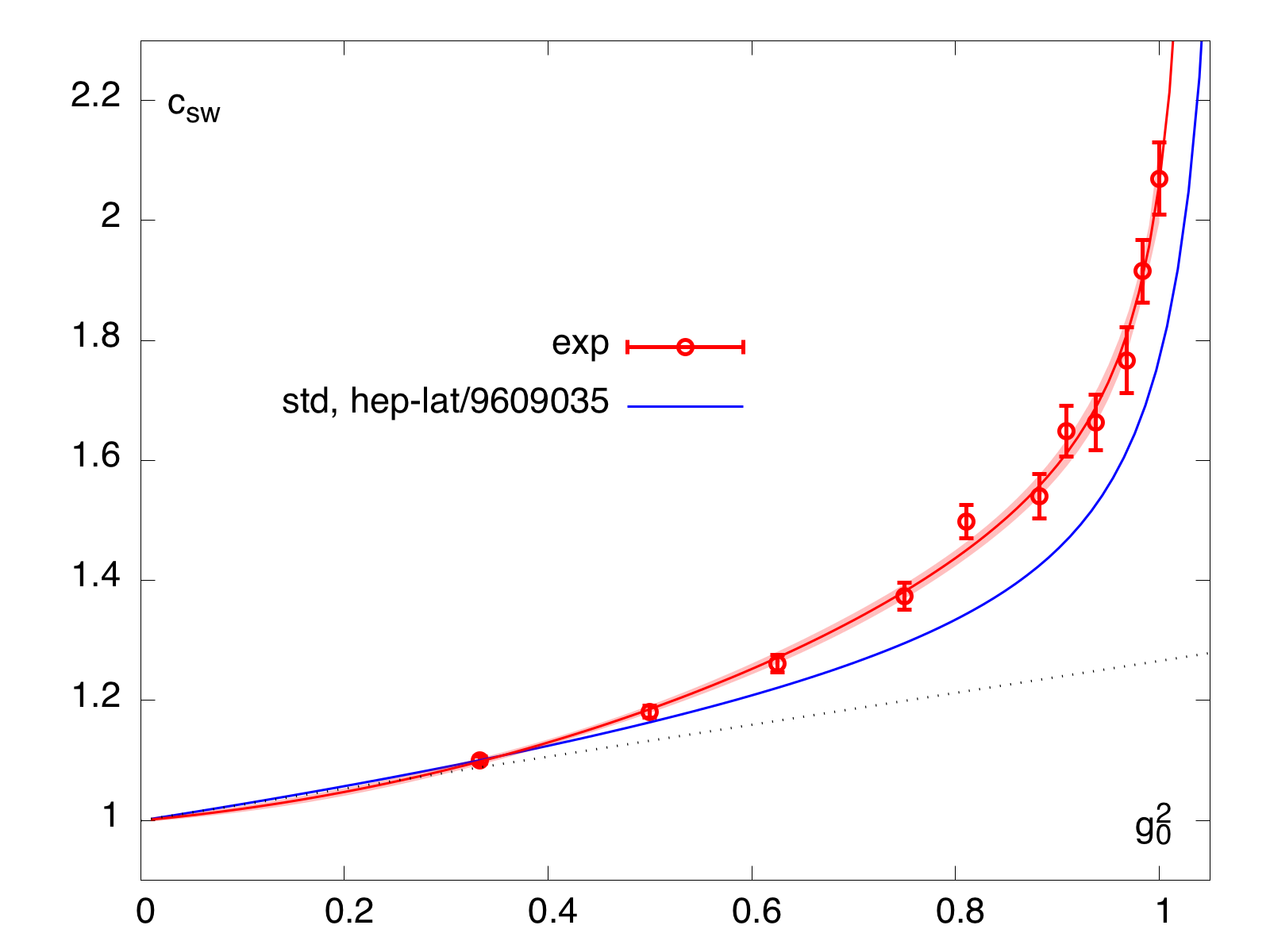}
\includegraphics[width=0.32\textwidth]{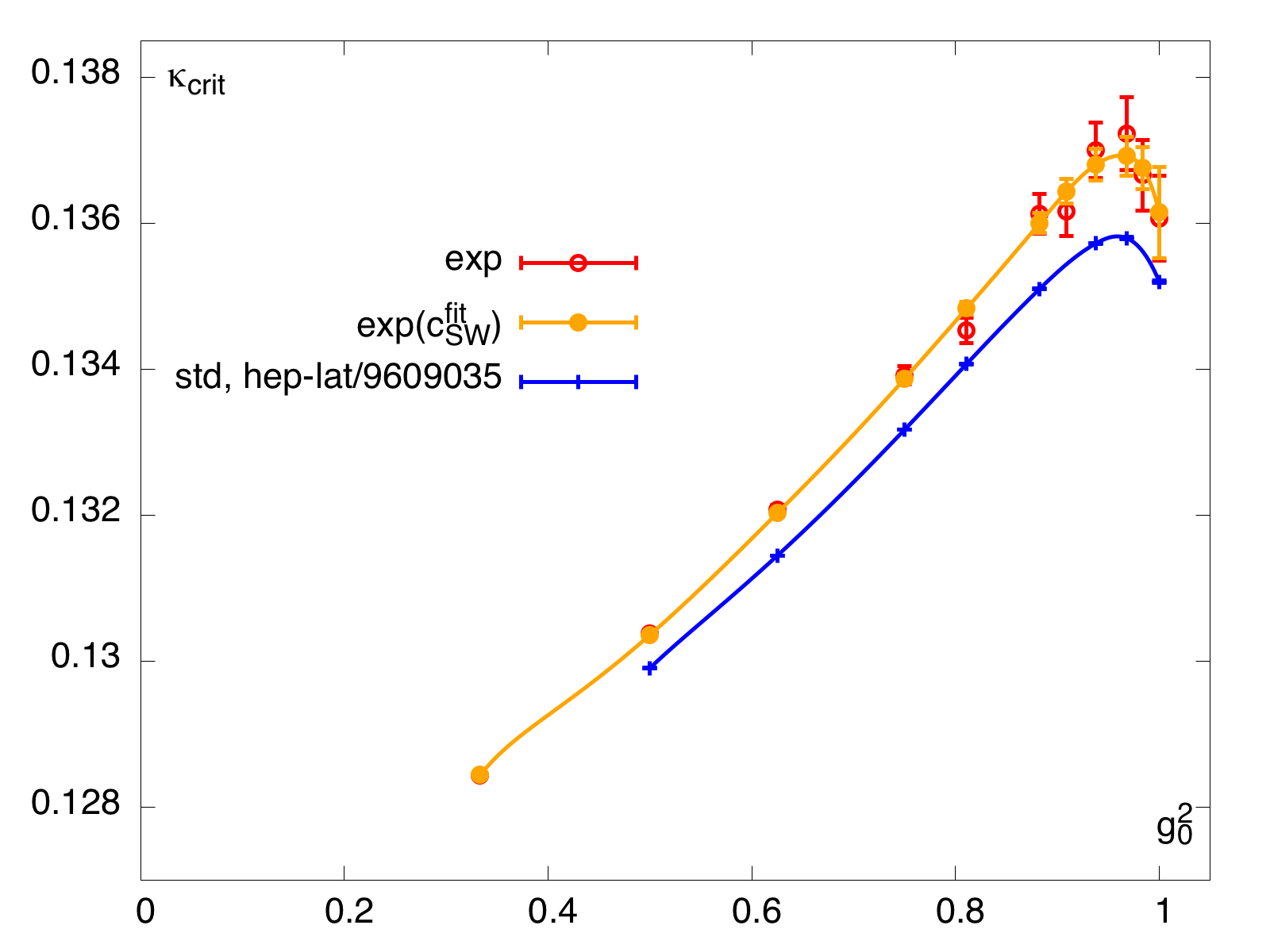}
\includegraphics[width=0.32\textwidth]{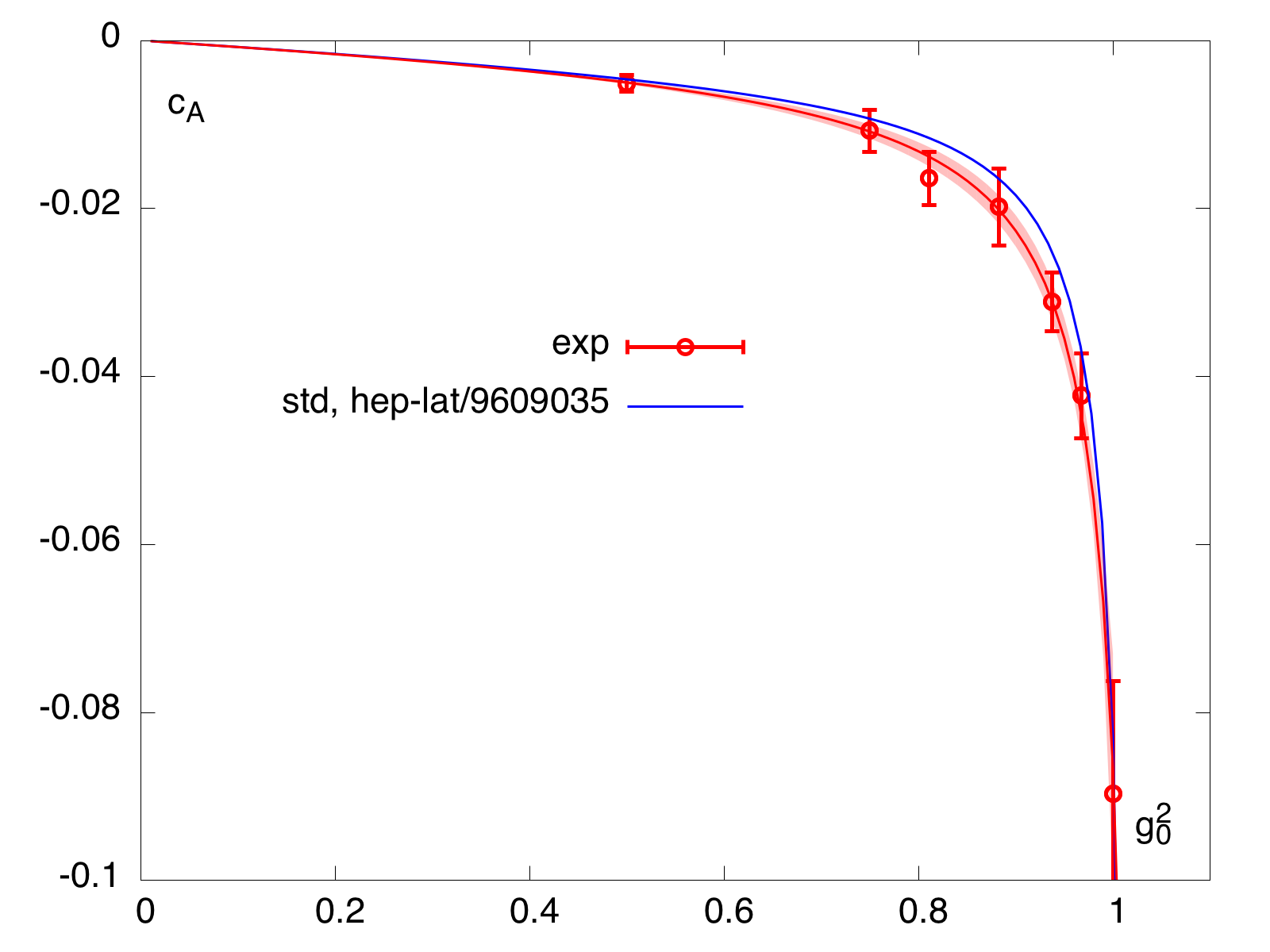}
\caption{\textit{
Non-perturbative tuning of $c_{SW}$ and $c_A$ as well as the measurement of $\kappa_\text{crit}$ in the SF in quenched QCD. We follow the tuning procedure outlined in \cite{Luscher:1996ug} and compare the standard clover (blue) from there with our exponentiated clover (red) results. Left: Determination of $c_{SW}(g_0^2)$.
Middle: Determination of $\kappa_\text{crit}$ using the numerical (red) and fitted (orange) results for $c_{SW}$.
Right: Determination of $c_{A}(g_0^2)$. 
}}
\label{fig:csw-kappa-ca}
\end{figure}

First, we determine $c_{SW}$ and $\kappa_\text{crit}$ by performing scans in $\kappa$ for a given $\beta$ in the range $\beta=6.0$ and $\beta=18.0$ and a number of trial $c_{SW}$ values. The results are linearly interpolated to find $M=0$, where $M$ is a particular definition of the unrenormalized current quark mass independent of $c_A$ as defined from the axial Ward-identity in the SF with background fields. This determines the point in $\kappa$ for the corresponding interpolation of $\Delta M$, where $\Delta M$ is a linear combination of Ward-identity quark masses that should vanish at finite lattice spacing when $c_{SW}$ is appropriately tuned, up to small tree-level corrections. Once $\Delta M(c_{SW})$ has been determined for a number of $c_{SW}$ values the results are again interpolated linearly to match the improvement condition:
$\Delta M=\Delta M^{(0)}
|_{M=0,c_{SW}=1}=0.000277/a$.
%
The determined values of $c_{SW}(g_0^2)$ are well described by:
\begin{equation}
c_{SW}(g_0^2)=\frac{1-0.7975 g_0^2 -(-0.2633) g_0^4-0.3675 g_0^6  }{1-0.9523 g_0^2 }~~~.
\label{eq:csw}
\end{equation}
The results for $c_{SW}(g_0^2)$ are plotted in Fig.~\ref{fig:csw-kappa-ca} (left). In all cases the blue band denotes the standard Clover results from \cite{Luscher:1996ug} while the red points and bands show those obtained using the exponentiated clover. 
Using the results for $c_{SW}$ - either direct numerical or from the interpolation - also $\kappa_\text{crit}$ can be determined.
Both are shown in Fig.~\ref{fig:csw-kappa-ca} (middle) in red and orange, respectively. The orange line denotes a cubic spline interpolation of $\kappa_\text{crit}(g_0^2)$.

For the improvement coefficient $c_A$ we next generate a number of SF-boundary condition ensembles without background field and calculate the required correlators with fermionic twists $\theta=1$ and $\theta=0$ as outlined in \cite{Luscher:1996ug}. This leads to the improvement condition:
$\Delta m=  m|_{\theta=1} - m|_{\theta=0} = \Delta m^{(0)}|_{m=0,c_A=0} = 0.000365/a$.
Performing tuning scans, $c_A$ can then be found in a similar way to $c_{SW}$.
The determined values for $c_{A}(g_0^2)$ are shown in Fig.~\ref{fig:csw-kappa-ca} (right). They can be described by the functional form:
\begin{equation}
c_A(g_0^2) = -0.00756 g_0^2 ~\frac{1-0.6347 g_0^2}{1-0.9680 g_0^2}~~.
\end{equation}

\subsection{Comparing the valence pion correlator for the standard and exponentiated clovers}

\begin{figure}[t!]
\centering
\includegraphics[height=0.22\textheight]{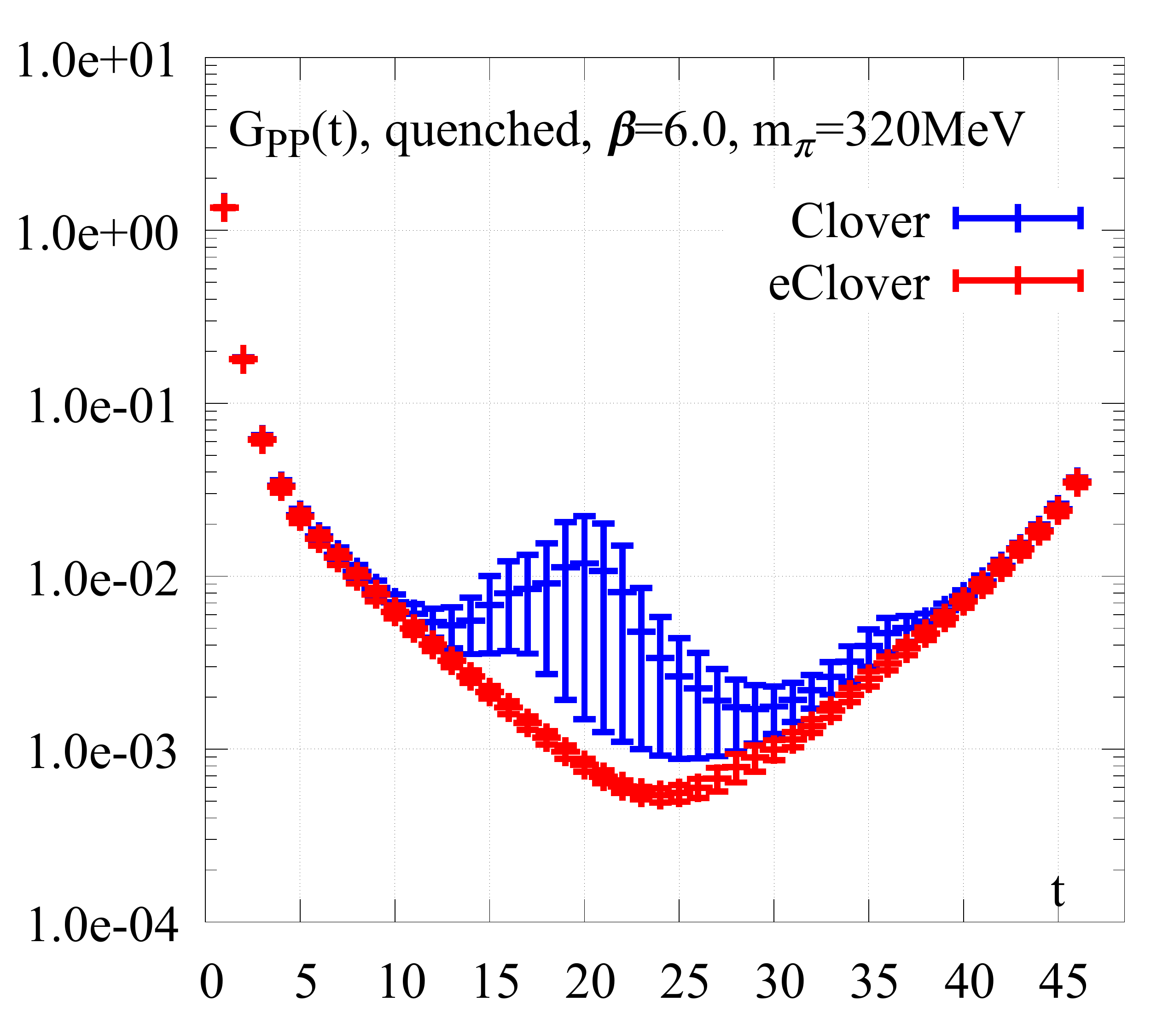}
\includegraphics[height=0.22\textheight]{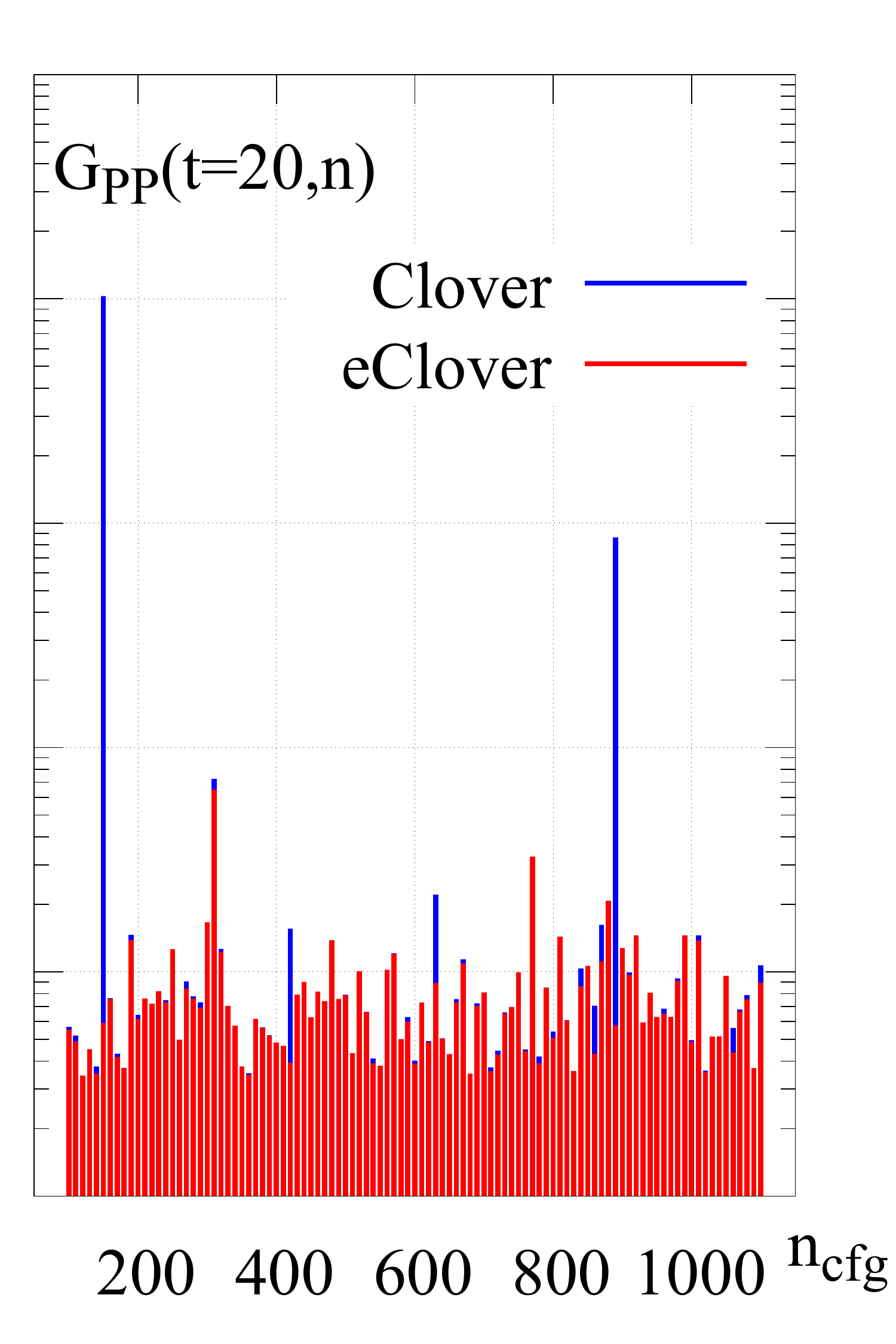}
\caption{\textit{ Left: The pion correlation function determined using the standard (blue) and exponentiated (red) clover terms. Right: The Monte Carlo time history of the pion correlator at Euclidean time $t/a=20$. Both correlators are calculated on the same gauge configurations with the same solvers using their respective non-perturbative improvement coefficients and tuned to the same mass.}}
\label{fig:quench-spec}
\end{figure}

To study the effect of the different clover terms on hadrons in the quenched theory, we compare the pion correlation function in a large volume calculation with periodic (anti-periodic for the valence quarks) boundary conditions. As first step we generated $N=1000$ independent configurations at $\beta=6.0$ ($a=0.093$fm) with $L^4=48^4$. On the same configurations we next calculate valence pion correlation functions using the non-perturbatively tuned improvement coefficients for both the standard and the exponentiated clover. We carefully tune $\kappa$ in both cases to achieve the same $PCAC$ quark masses. 
Our aim is to compare the performance of both clover terms in a regime where quenched calculations are known to suffer from the frequent occurrence of so-called exceptional configurations. As such we set the valence masses to $m_{\pi}^\text{val}=320$~MeV, whereby we use the same solver and parameters in both cases. To convert to physical units we use the scale set in \cite{Giusti:2018cmp}. 
The comparison is given in Fig.~\ref{fig:quench-spec}, with the standard clover in blue and the exponentiated clover in red. The left panel shows the bootstrapped pion correlation functions, while the right panel shows the Monte Carlo time history of the correlators at Euclidean time $t/a=20$. We observe that the exponentiated clover does not exhibit any exceptional configurations, as the large spikes observed in the standard clover in the Monte Carlo history are absent. The correlation function is well behaved for all distances, while for the standard clover exceptional configurations dominate the signal in the long distance regime. Recall that these measurements are on identical configurations.

\section{Dynamical QCD with $2+1$ quark flavours}

Moving forward from the quenched case we next address full QCD with a light isospin doublet pair of quarks and a single strange quark, albeit the mass of the strange quark is not set to its physical value, as explained in Sec.~\ref{sec:trajectory}. 
In the following we describe our overall setup and tuning strategies as well as tuning goals. The results shown comprise some published in \cite{Francis:2019muy} in addition to new ones.
Throughout, to set the scale we use the gradient flow time criterion and 
to convert our calculations into physical units we employ the gradient flow time 
$\sqrt{t_0}=0.1464(18)\textrm{fm}$, corresponding to 
$\sqrt{8t_0}=0.414(5)\textrm{fm}$ taken from \cite{Bruno:2016plf}.

\subsection{Non-perturbative tuning of $c_{SW}$ - extending the reach in $a$}

\begin{figure}[t!]
\centering

\includegraphics[width=0.28\textwidth]{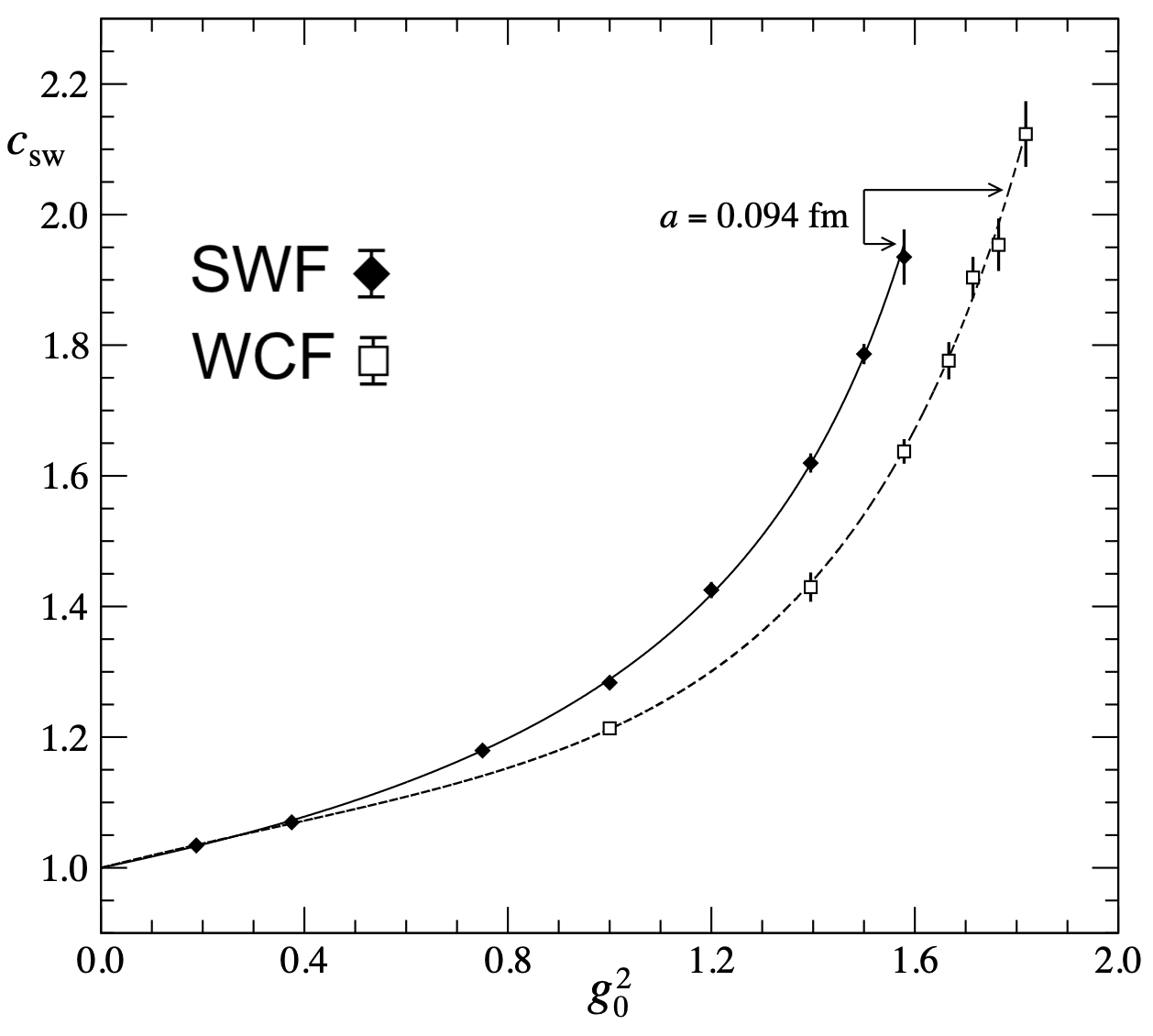}
\includegraphics[width=0.34\textwidth]{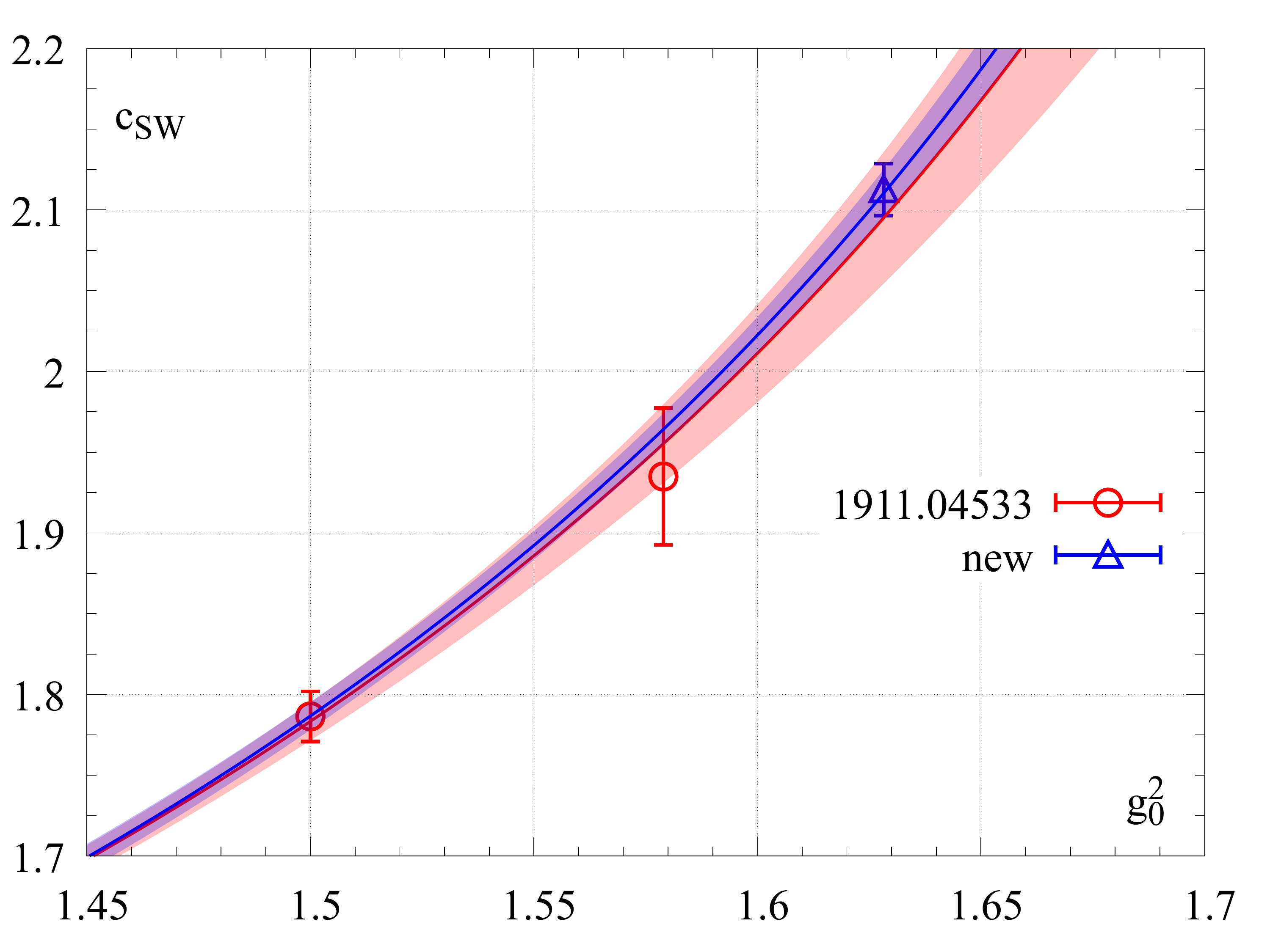}
\includegraphics[width=0.34\textwidth]{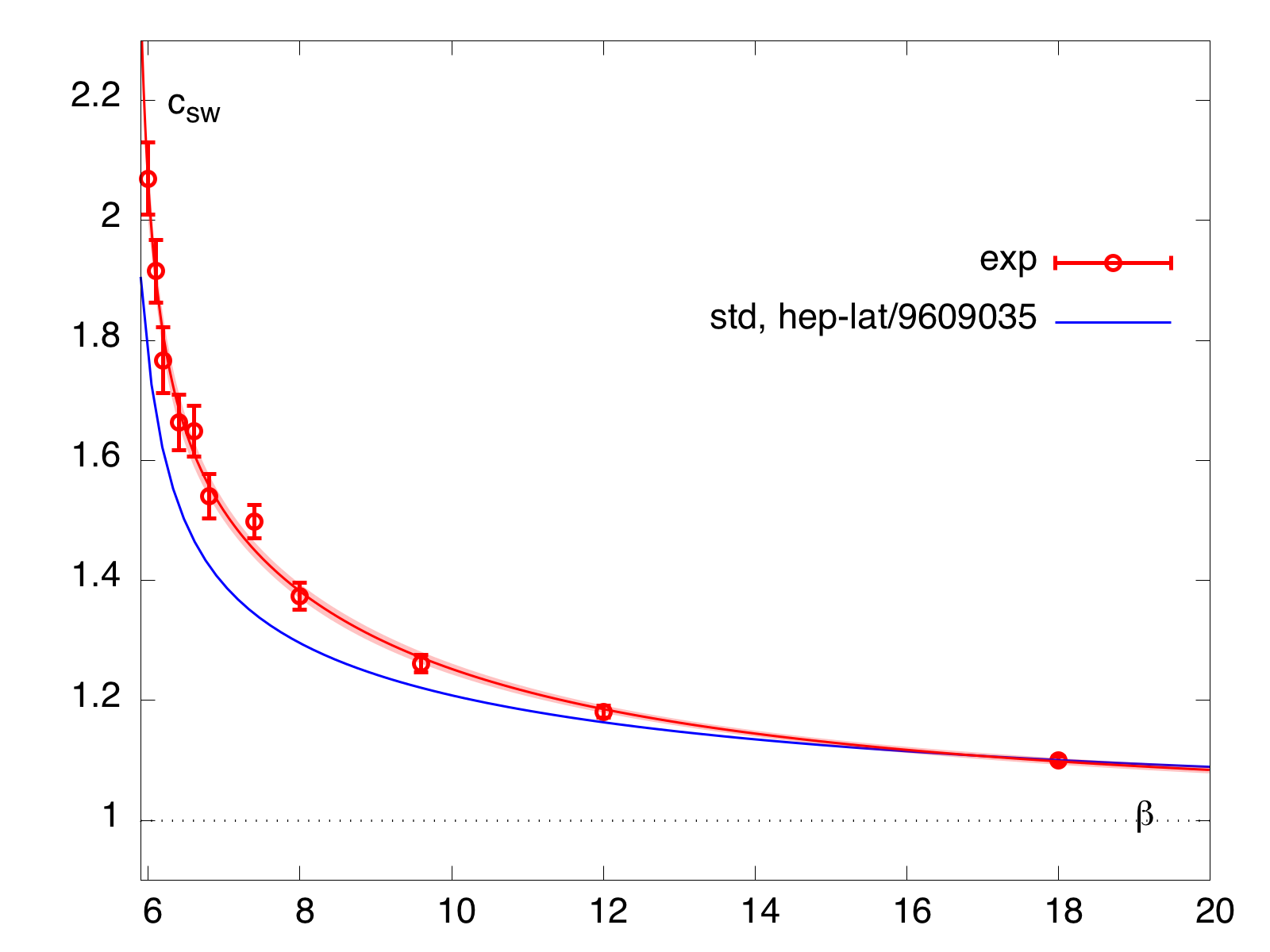}
\caption{\textit{Non-perturbative tuning of $c_{SW}$ using the Schr\"odinger functional for $N_f=3$ full QCD. Left: In \cite{Francis:2019muy} we compared the results of the SWF with those using the standard WCF setup \cite{Bulava:2013cta}. Middle and right: Extending the range in $g_0^2$ we tune $c_{SW}$ also for $a=0.12$~fm. The red band denotes the interpolation formula from before while the blue shows the updated result.}}
\label{fig:dyn-csw}
\end{figure}

Turning to full QCD the clover coefficient must once more be fixed and in \cite{Francis:2019muy} the procedure of \cite{Luscher:1996ug} was used to perform the tuning in the range $\beta\geq 3.8$ in $N_f=3$ QCD. There, simulations with three, mass-degenerate, dynamical flavours with $SF$ boundary conditions where performed in small volumes of $L=8$ and $T=16$. A similar procedure was followed for the standard WCF setup in \cite{Bulava:2013cta}. The resulting comparison figure is shown once more for reference in Fig.~\ref{fig:dyn-csw} (left). 
%

Here, we extend the range of the non-perturbative tuning to $\beta\geq 3.685$. This time a larger volume of $L=16$ and $T=16$ was used. The choice of $\beta=3.685$ corresponds to $a=0.12$~fm (determined via $t_0/a^2$). Such a coarse lattice spacing has interesting applications for methods development and nuclear physics studies in particular.
Adding this new point to the previous ones we obtain the results shown in Fig.~\ref{fig:dyn-csw} (middle, right), where the blue band denotes a re-interpolation of the clover coefficient. We find good agreement, within $1\sigma$, with the previously determined interpolation formula and confirm
\begin{equation}
c_{sw}(g_0^2) = \frac{1-0.325022g_0^2-0.0167274g_0^4}{1-0.489157g_0^2}~~
\end{equation}
in the extended range $\beta=6/g_0^2\geq 3.685$.

\subsection{Setting the chiral trajectory}
\label{sec:trajectory}

Renormalisation and $\mathcal{O}(a)$-improvement complicate the approach to the continuum limit for certain observables with Wilson fermions. As a result special care needs to be taken when choosing the trajectory in bare parameter space to physical masses and the continuum limit. Here, we choose a tuning strategy applied previously by the QCDSF~\cite{Bietenholz:2010jr} and CLS~\cite{Bruno:2014jqa} collaborations.

The idea is to simplify tuning of the mass-parameters at given gauge coupling and reducing mass-dependent cutoff effects by working at constant trace of the simulated quark mass matrix ($\textrm{tr}[M]=\textrm{const}$). The tuning starts at the flavour-symmetric point where all pseudoscalar meson masses are degenerate and
$    \textrm{tr}[M] = m_u + m_d + m_s = N_f \cdot m_\ell = \textrm{const} $
depends on a single mass parameter. As physical input the ground state masses of the pion and kaon are chosen, and tuning $m_\ell$ amounts to matching the meson mass combination 
\begin{equation}
    m_{\pi K}^2 = \frac{2}{3}\Big( m_K^2 + m_\pi^2/2\Big) \equiv m_{\pi K}^2|_{\textrm{phys}}
\end{equation}
to its physical value. This definition is motivated by leading order chiral perturbation theory, from which it is known that $m_{\pi K}^2\propto \textrm{tr}[M]$. Corrections due to higher order contributions are known to be reasonably small. With the starting point tuned, the masses between strange and the degenerate light isospin doublet $m_s\neq m_u = m_d$ is split by keeping  $\textrm{tr}[M]=\textrm{const}$. When the pion mass is decreased to its physical value, the kaon mass also approaches its physical value as a result. 

When choosing the input for matching the lattice theory to experiment, we have to subtract small (perturbative) effects arising from strong isospin breaking and electromagnetic effects in physical hadron masses. We follow the procedure outlined in \cite{Aoki:2016frl}, leading to input values (in MeV)
\begin{equation}
    m_\pi|_{\textrm{phys}}=134.8(3),~~m_K|_{\textrm{phys}}=494.2(3)~~\Rightarrow~~m_{\pi K}|_{\textrm{phys}}=410.9(2)~. 
\end{equation}
Similarly, we quote the physical values for the pion and kaon decay constants (in MeV)
\begin{equation}
    f_\pi|_{\textrm{phys}}=130.4(2),~~f_K|_{\textrm{phys}}=156.2(7)~~\Rightarrow~~f_{\pi K}|_{\textrm{phys}}=147.6(5)~.
\end{equation}

\subsection{Validation}
\label{sec:valid}

Once the above outlined trajectory is tuned using the already available non-perturbatively $c_{SW}$, the generation of large volume gauge ensembles with well defined physical mass and continuum limits is enabled.
To ensure the trajectory is followed to sufficient accuracy for the argument to hold as well as to ensure the overall validity and quality of the gauge configurations produced, a number of tuning goals and quality criteria should be fixed. 
%
As the understanding of the used algorithms and QCD itself increases this list is constantly extended.
Here, we highlight a small number of examples of observables aimed at showing stability of the generation process. To label a set of configurations as safe we set the targets:
\begin{itemize}
\itemsep0em
    \item For the chiral trajectory the tuning quantity $\phi_4=8t_0( m_K^2 + m_\pi^2/2)$ is within $0.5\%$ of the target value of $1.115$ with an error of max. $1\sigma$. 
    \item Total reweighting factor fluctuations are mild and ideally below $5\%$. 
    \item The SMD step distance $\delta \tau$ is set to maximise the backtracking period $t_{acc}=\delta \tau\, P_{acc}/(1 - P_{acc})$ \cite{Luscher:2011kk}.
    \item Distribution of $\delta H$ matches that set by the acceptance rate. 
    \item Distribution of the lower and upper bounds of the spectral gap for the strange quark are within the input ranges and the degree of the Zolotarev is sufficiently high, $12 (V/2) \delta^2 < 10^{-4}$ \cite{mluscher:openqcd}.
    \item Observed well-behaved and gapped distribution of the lowest Dirac operator eigenvalue.

\end{itemize}
Furthermore, we carefully estimate the distance between two configurations, labelled as independent, based on the autocorrelation time of the topological charge $Q$ computed via the gradient flow.  
The boundary conditions are changed to open boundaries once we observe a marked and significant increase in the autocorrelation time of $Q$ signalling a possible freezing of topology. 
We check that there are no visible thermalisation effects in all of the above as well as in the plaquette, $t^2\langle E\rangle$ as well as $PP$ and $PA$ correlators in addition to observing the rule-of-thumb of thermalising a minimum of five autocorrelation lengths.

\section{First Ensembles and preliminary results}

With some of the overall features of the SWF setup established in the preceding studies we study their scaling properties in large (non-SF) volumes for a few observables. 
In particular we focus on the ensembles listed in Tab.~\ref{tab:used-gauges}. We adopt a naming convention in which simplified values of the lattice spacing $a$ and the pion mass $m_\pi$ are combined into a unique label.

\begin{table}[t]
\begin{center}
\begin{tabular}{c|cccccc}
\hline
\hline
  label & $a$ (fm) & $m_{\pi}$ (MeV) &$\beta$ & $\kappa_{ud}$ & $\kappa_s$ & dimension \\
\hline
  a12m400 & 0.12 & 410 & 3.685  & 0.1394400 & 0.1394400 & $96\times 24^3$  \\\hline
	a094m400 & 0.094 & 408 & 3.8  & 0.1389630 & 0.1389630 & $96\times 32^3$ \\
	a094m300 & 0.094 & 293 & 3.8  & 0.1391874 & 0.1385164 &  $96\times 32^3$ \\
	a094m200S & 0.094 & 215 & 3.8  & 0.1392888 & 0.1383160 &  $96\times 32^3$ \\\hline
	a064m400 & 0.064 & 409 & 4.0  & 0.1382720 & 0.1382720 &  $96\times 48^3$  \\\hline
	a055m400 & 0.055 & 412 & 4.1  & 0.1379450 & 0.1379450 &  $96\times 48^3$  \\
\hline
\hline
\end{tabular}
\caption{\textit{Details of the gauge field ensembles used for the scaling studies and determination of the hadron spectrum shown below. A listing of all available configurations is not given as we are currently producing more. The subscript added to {\tt a094m200S} denotes the small volume of the ensemble, we are currently producing a larger one as discussed in Sec.~\ref{sec:openlat}.}}
\label{tab:used-gauges}
\end{center}
\end{table}

There are some differences in the methodologies and statistics quoted for the the different analysis performed.
For the study of the continuum limit at the SU($3$)$_F$ symmetric point, we have analyzed $100$ independent configurations for each of the $4$ lattice spacing. The list of observables contains the simplest hadrons and the analysis method follows Ref.~\cite{Bruno:2014jqa} 
connecting well to our previous work in \cite{Francis:2019muy}.
To study the pion mass dependence we have analyzed a larger number of gauge configurations and enlarged the number of observables, including the baryon decuplet and full octet spectrum. Also a more advanced analysis methodology is deployed.
This reflects our growing data repository and evolving analysis methodologies as more varied and complex observables become available. All results presented here are preliminary.

\subsection{First scaling studies and initial continuum extrapolations at the $SU(3)_F$ point}

\begin{figure}[t!]
\centering
\includegraphics[width=0.47\textwidth]{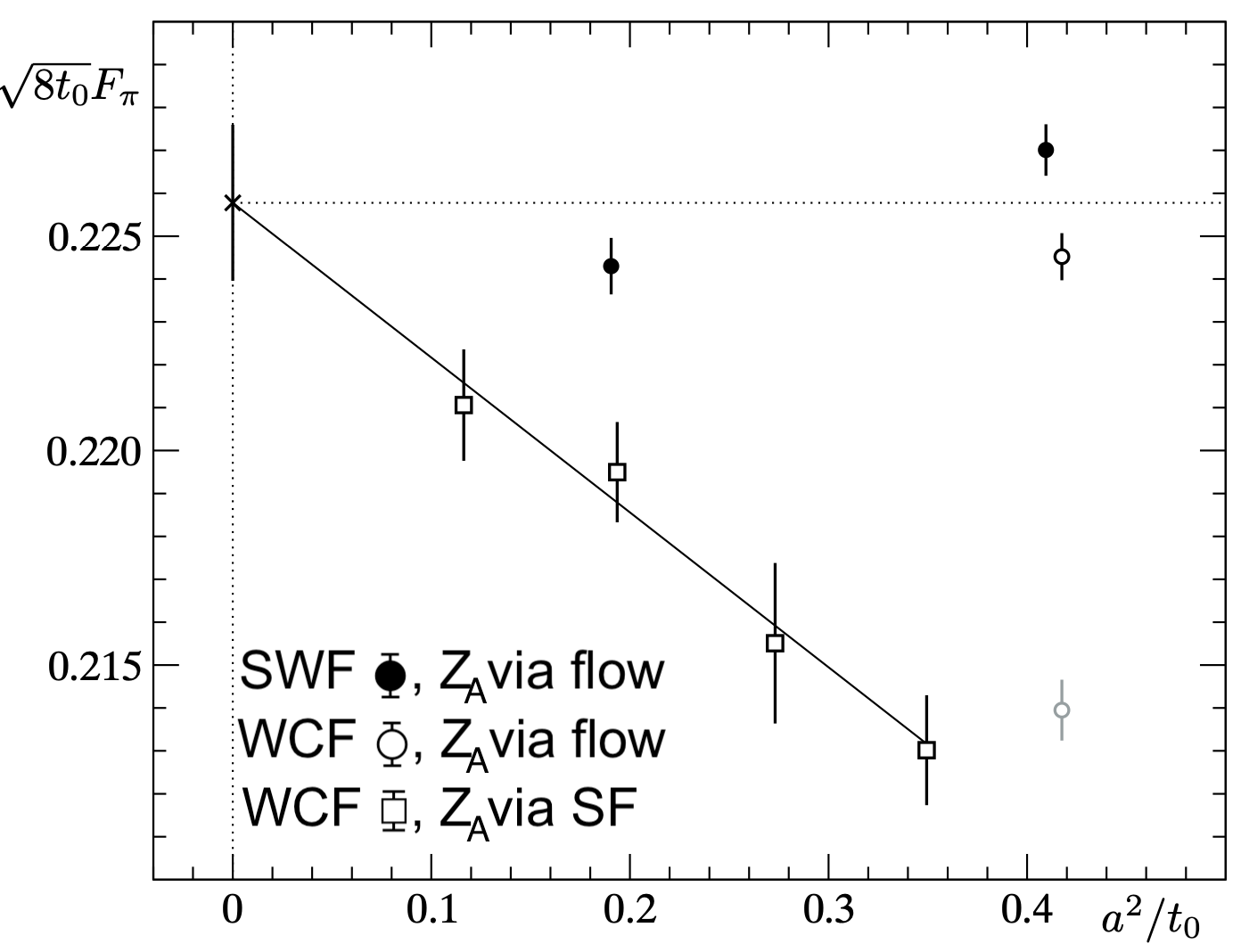}
\includegraphics[width=0.52\textwidth]{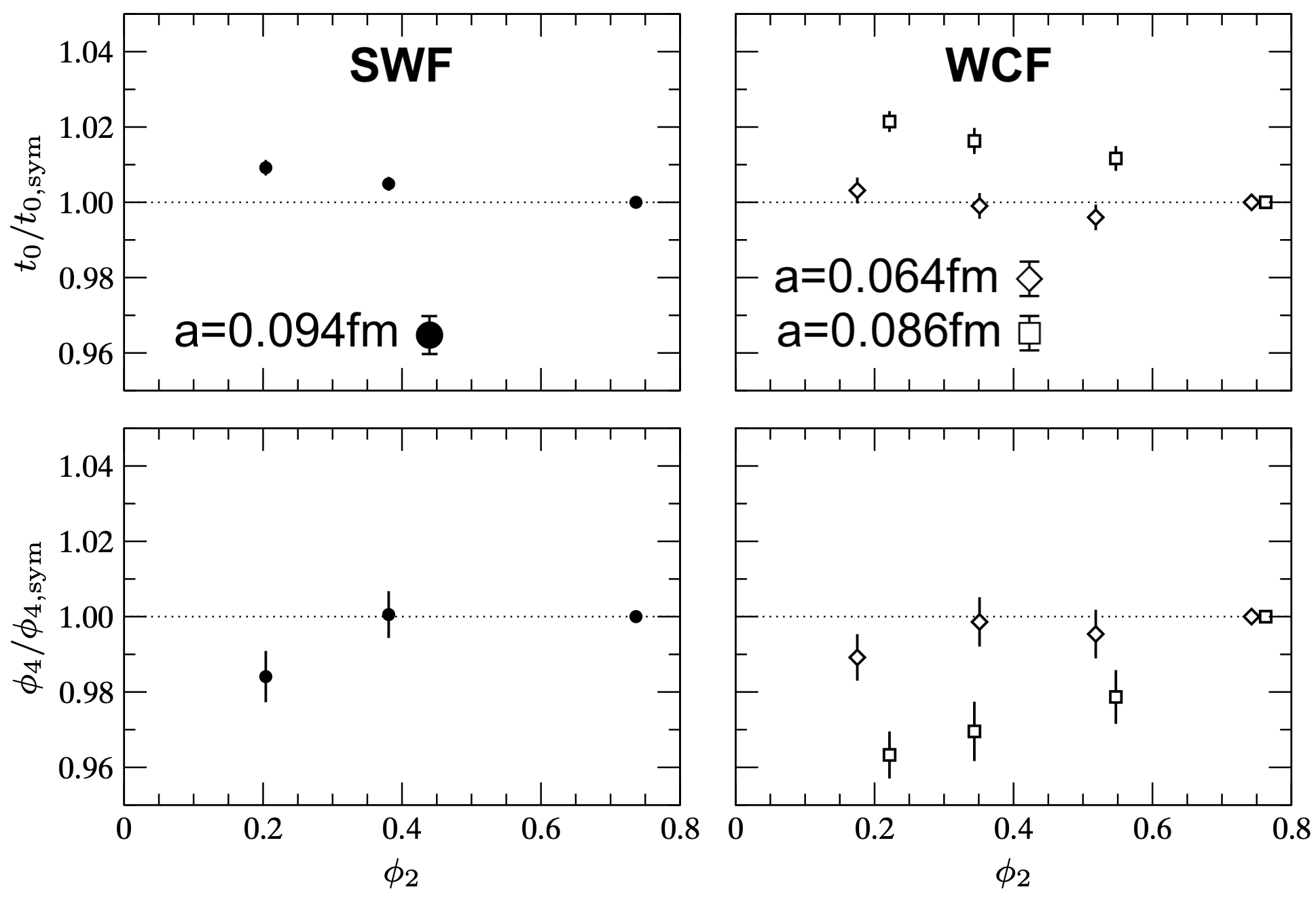}
\caption{\textit{First scaling studies at $a=0.094$ and $0.064$~fm.
Left: $f_\pi$, renormalised using the gradient flow, scaling in lattice spacing $a$. Right: Chiral scaling at fixed $a$ for the quantities $t_0/t_{0,\rm{sym}}$ (top) and $\phi_4/\phi_{4,\rm{sym}}$ (bottom) with SWF (middle) and WCF (right) setups. Figures from \cite{Francis:2019muy}. }}
\label{fig:cont-old}
\end{figure}

To start our study of the scaling properties of the SWF setup, we show the approach to the continuum of the renormalised pion decay constant $f_\pi$ at the flavour symmetric point in Fig.~\ref{fig:cont-old} (left) from \cite{Francis:2019muy}. The full symbols denote results obtained using SWF on a094m500 and a064m400, while the open squares show results obtained using WCF \cite{Bruno:2016plf}. A key difference here is the renormalisation, as in the latter $Z_A$ was computed in the SF scheme, while in the SWF case we chose to determine $Z_A$ using the gradient flow~\cite{Luscher:2013cpa}. This has advantages for the scaling but also lets us eliminate the effect of $c_A$ \cite{Luscher:2013cpa,Francis:2019muy}. However, this obscures the continuum scaling and therefore we added simulations with WCF renormalised in the same way as the SWF (open circles).

Turning to relative $am_q$ effects, in Fig.~\ref{fig:cont-old} (middle and right) we compare ratios of observables at the flavor symmetric point and their counterparts at lower quark mass. In particular we study $t_0/t_{0,\rm{sym}}$ and $\phi_4/\phi_{4,\rm{sym}}$ for the SWF (middle) and the WCF (right). The latter were computed on configurations generated and presented in \cite{Bruno:2014jqa}. The SWF results were derived from the ensembles a094m400, a094m300 and a094m200. The $x$-axis in these cases is labelled by $\phi_2=m_\pi^2 8t_0$. These results clearly indicate that, on the lines of constant physics chosen here and for these $2$ observables, $t_0$ and $\phi_4$, cutoff effects for SWF 
are reduced in comparison with WCF.

\begin{figure}[t!]
\centering
\includegraphics[height=0.24\textheight]{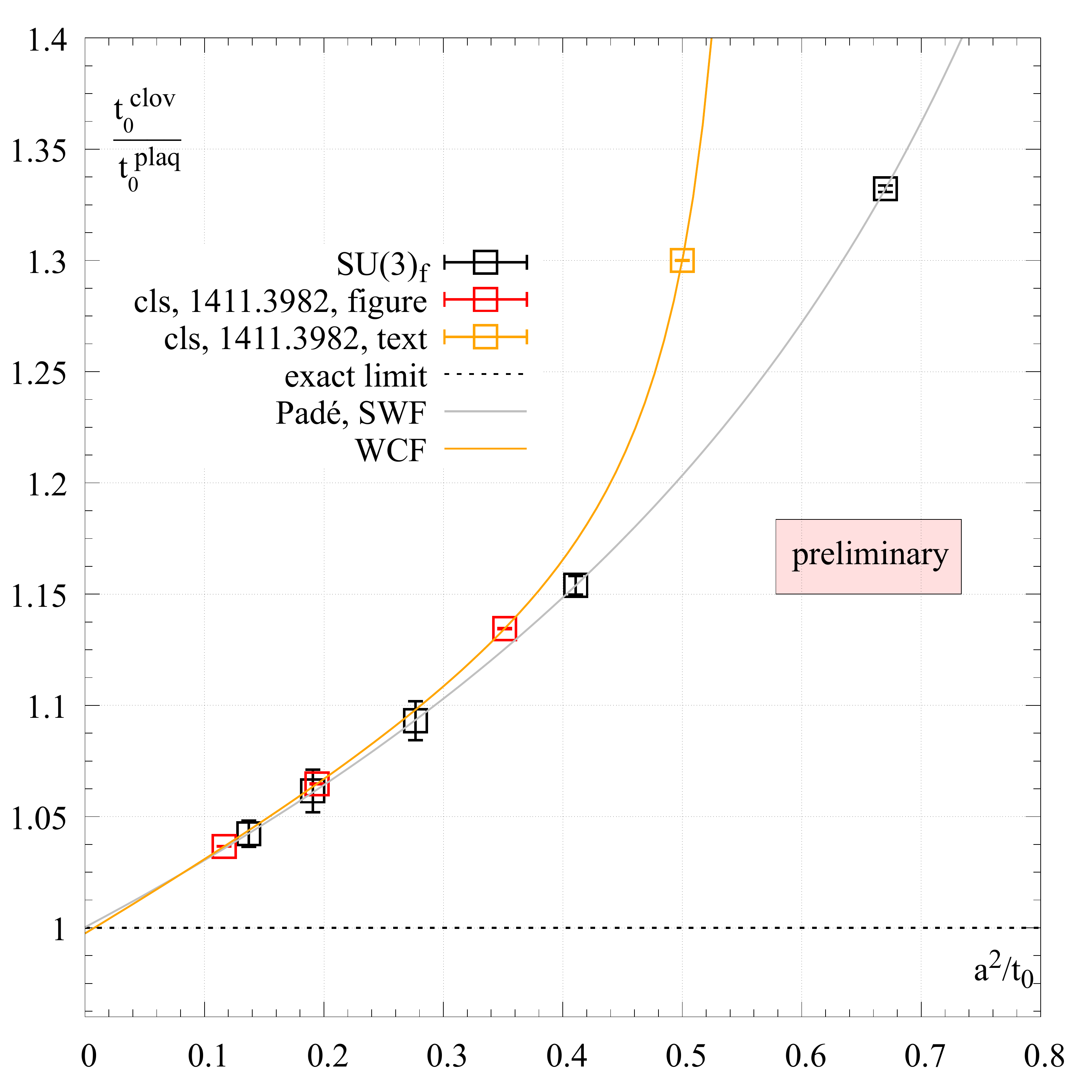}
\includegraphics[height=0.24\textheight]{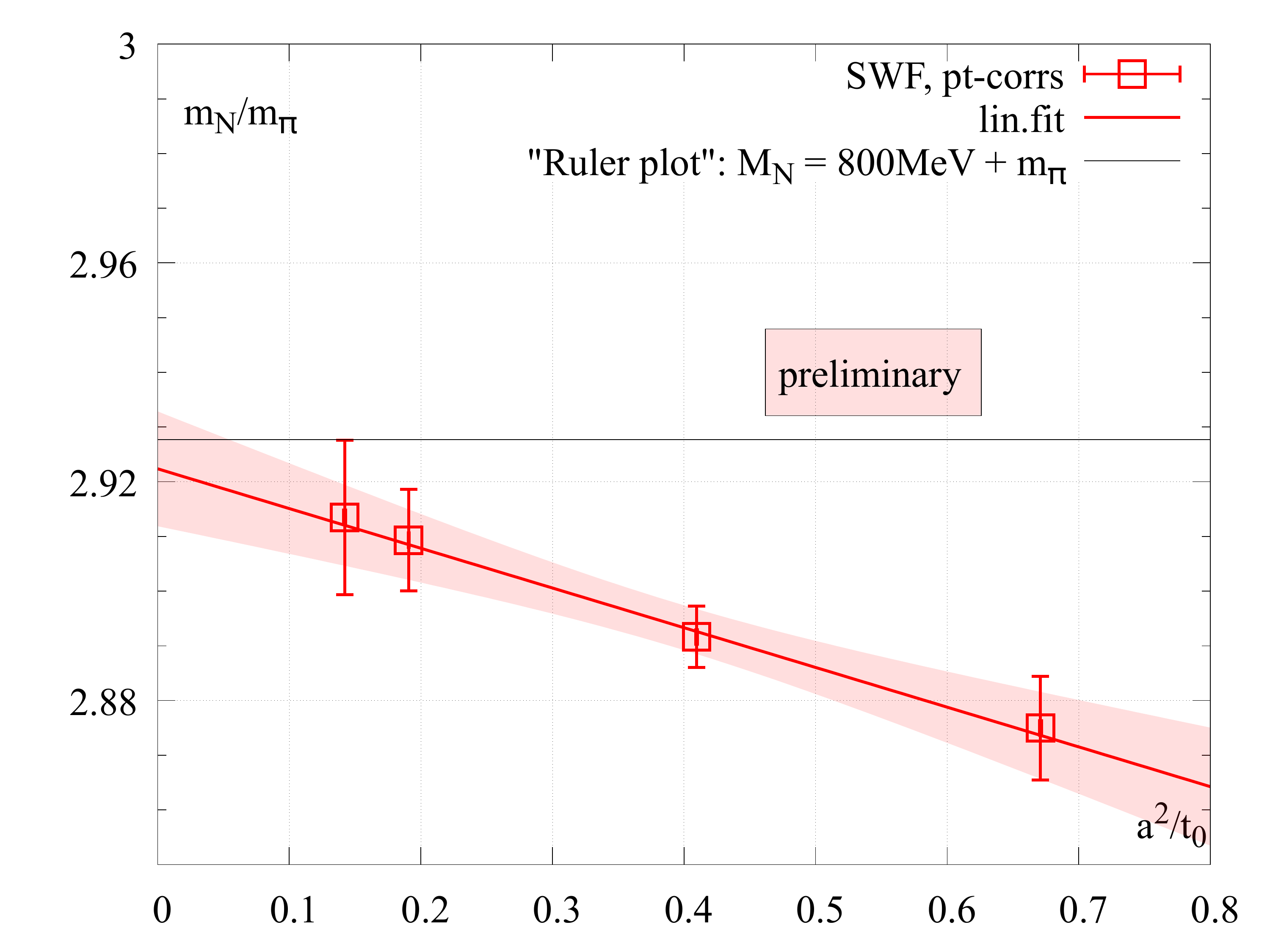}
\caption{\textit{First scaling studies including new, preliminary, ensembles.
Relative effects in $t_0^{\rm{clov}}/t_0^{\rm{plaq}}$ (left) and $m_{nucleon}/m_{\pi}$ (right) show good scaling in the SWF setup. These observations need to be further confirmed before firm conclusions can be drawn.}}
\label{fig:cont-new}
\end{figure}

Adding to these previously known results we next turn to new results using also our new ensmebles. At the SU($3$)$_F$ flavor-symmetric points we now have access to four lattice spacings $a=0.12, 0.094, 0.064, 0.055$~fm. Note, that the last uses open boundary conditions, unlike the other ensembles, as we observed a significant slow down in topological charge tunneling. 
On these new ensembles we first determined the ratio $t_0^{\rm{clov}}/t_0^{\rm{plaq}}$. The continuum limit of this quantity is exactly 1 and it was studied using WCF \cite{Bruno:2014jqa}, giving an interesting opportunity for a comparison. The results are shown in Fig.~\ref{fig:cont-new} (left) as a function of $a^2/t_0$, whereby the black squares denote the SWF and the red/orange the WCF results obtained from \cite{Bruno:2014jqa}. 
We fit both data sets with a Pad\'e ansatz and we find that higher order discretization effects set in at finer lattice spacings for WCF.
Since $m_\pi$ is used to tune the trajectory and set the flavour symmetric point, the next stable particle that we can study with predictive power is the nucleon. Using the same ensembles as in the preceding study of $t_0^{\rm{clov}}/t_0^{\rm{plaq}}$, we determined the nucleon correlation function using point sources with $100$ configurations and $16$ stochastic source locations. We then determined the nucleon mass by performing $2$-state and $1$-state fits to determine the longest stable plateaus following the procedure outlined in \cite{Bruno:2014jqa} and quote the ground state mass and error of the $1$-state fits.
Plotting the results as $m_N/m_\pi$ vs. $a^2/t_0$ in Fig.~\ref{fig:cont-new} (right) we perform a rudimentary continuum extrapolation with a linear fit in $a^2$. 
At the level of accuracy achieved we observe good scaling properties up to $a=0.12$~fm in the nucleon-to-pion mass ratio. For reference we give the ``ruler plot'' result \cite{Walker-Loud:2008rui,Walker-Loud:2014iea} at this pion mass as well.
Both studies are still at a preliminary level and the nucleon mass in particular should be considered at an early stage. 
We emphasize that the indications of reduced cutoff effect for SWF, in comparison with WCF, we found in these preliminary studies refer to 
the specific observables analyzed here. Whether this is a general feature of this new lattice action is left to future studies.
If further confirmed, the good scaling in $m_N$ up to $a=0.12$~fm at the flavour symmetric point could be particularly interesting. It can directly impact nuclear physics applications, e.g. the H-dibaryon \cite{NPLQCD:2010ocs,Inoue:2010es,Francis:2018qch}, where larger than expected discretisation effects have been shown to have the potential to spoil the conclusions drawn from finite-$a$ calculations once the continuum limit \cite{Green:2021qol,Green:2021sxb} is taken.

\subsection{First results on the pion mass dependences of hadrons}

In this section we present a spectroscopic analysis of light and strange hadron masses on SWF gauge ensembles with the goal to establish a first look at their pion mass dependence. The calculation of the hadron spectrum has been performed using the \texttt{lalibe} \cite{callat:lalibe} software package, built on top of \texttt{Chroma} \cite{Edwards:2004sx,Clark:2009wm}. The codes have been modified to include the exponentiated clover term needed for SWF simulations, both for CPU and for GPU calculations. 
In the following, we select the ensembles a094m400, a094m300 and a094m200S, 
with $N=234, 167, 210$ evaluated configurations, respectively. 
We used point and smeared sources, with $16$ source positions each, for the correlation functions. We fix the smearing setup by studying four sets of gauge invariant smearing parameters, $N^\text{smear}$ and $\sigma$, the input parameters of the \texttt{GAUGE\_INV\_GAUSSIAN}
routine in \texttt{Chroma}, as suggested in \cite{Gusken:1990smear}. The parameters with the best trade-off between the reduction of excited state contamination and the loss of signal are chosen for the following analyses. We find that $N^\text{smear}=32$ for the number of iterations and $\sigma=3.86$ for the smearing width are a good choice. 

\begin{figure}[t]
    \centering
    \includegraphics[width=0.43\textwidth]{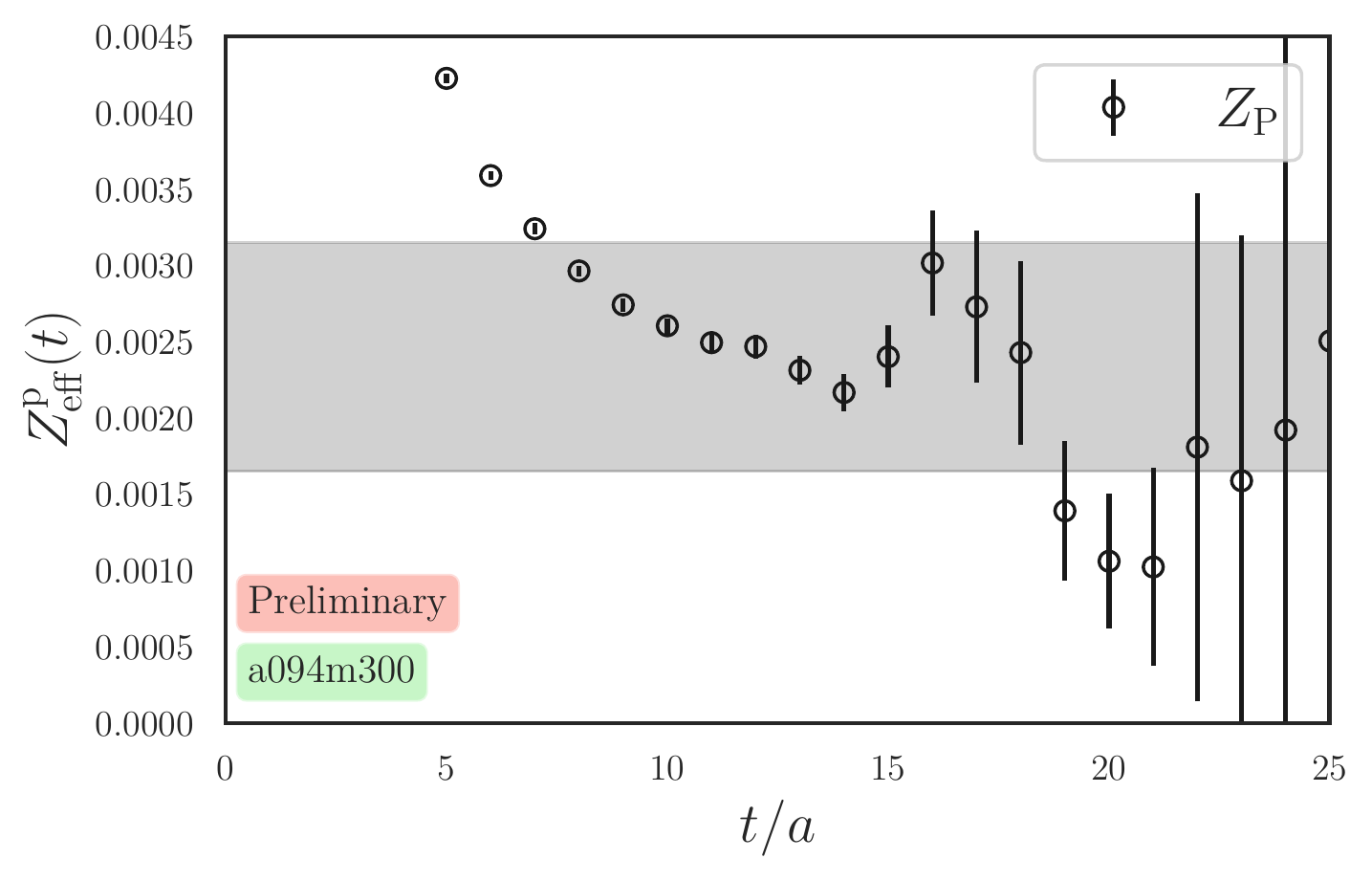}
    \includegraphics[width=0.43\textwidth]{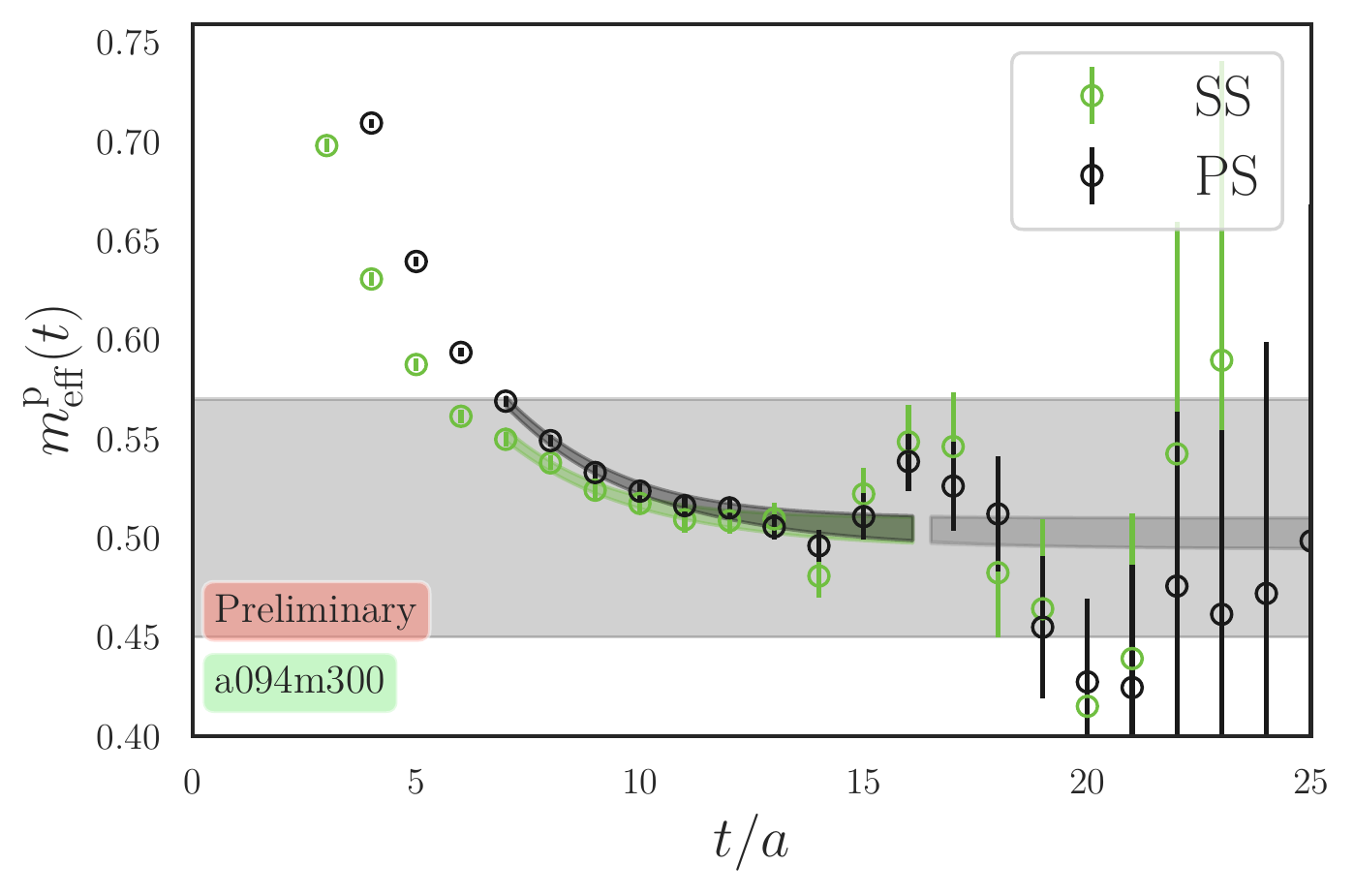}
    \includegraphics[width=0.43\textwidth]{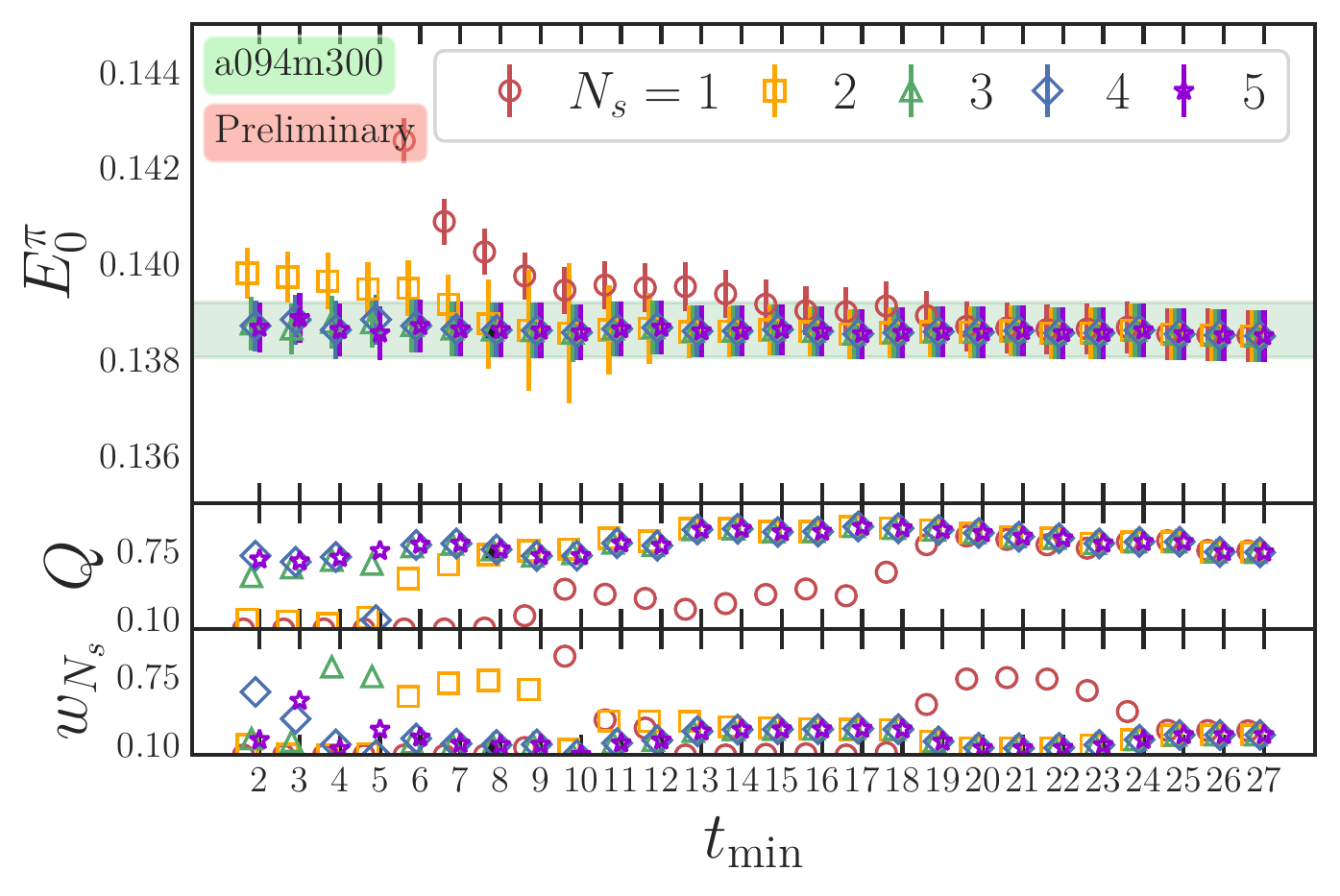}
    \includegraphics[width=0.43\textwidth]{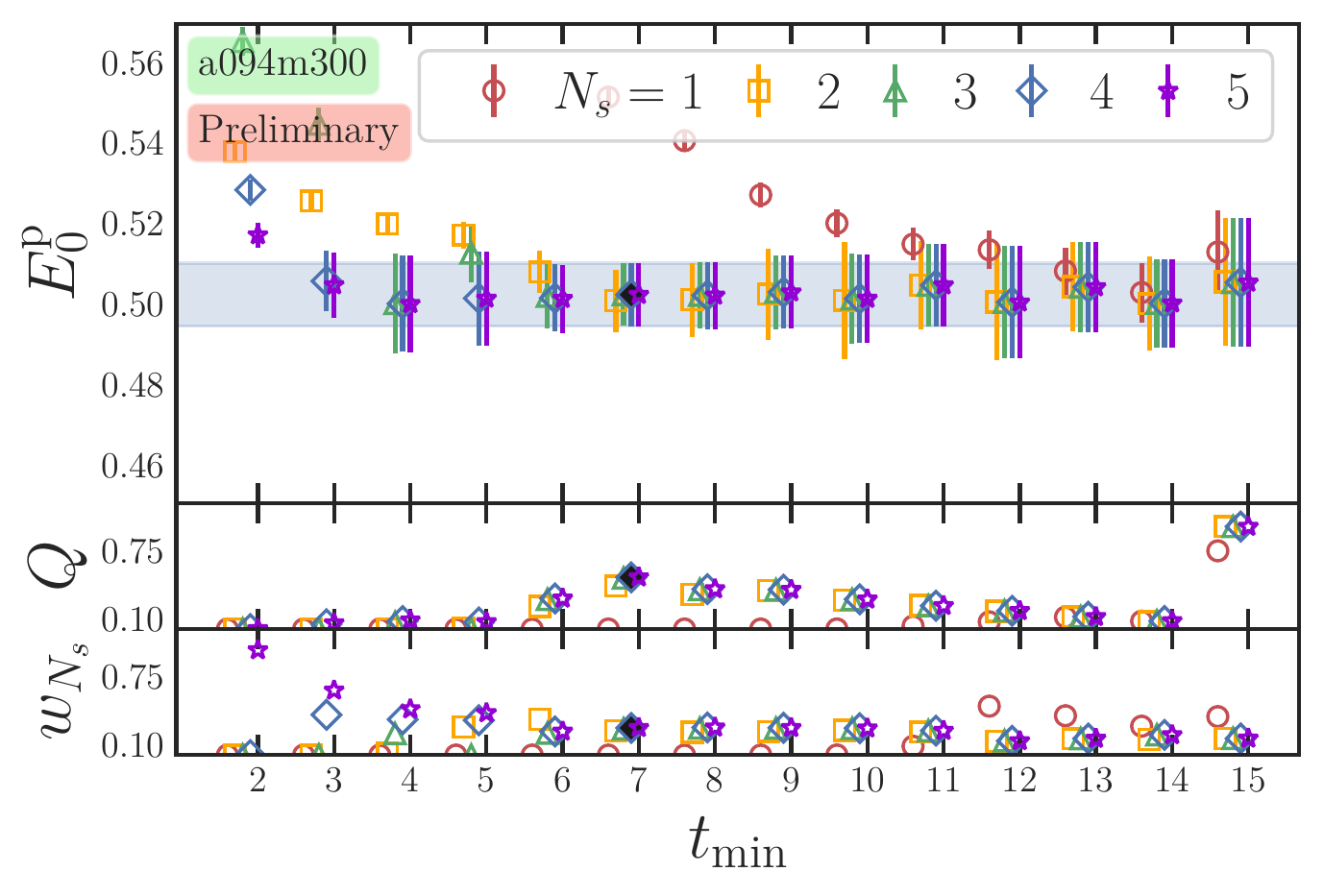}
    \caption{\textit{Top row: Amplitudes (left) and effective masses (right) in lattice units for the proton on a094m300. In the left panel we show the SP results only. Here, the gray bands represent the prior width of $1\sigma$. The dark gray band on the right represents the selected fitted time window. An extrapolation beyond this range is shown after a break in the fit band. Bottom row: Fit window dependence of our results for the pion (left) and proton (right). The figures show (from top to bottom): The value of $E_0(t_{min},N_s)$, the corresponding $Q$ values and the relative weights, based on the Bayes factor. The chosen fit window and $E_0$ are denoted by a filled black symbol, $t_{min}=8$ (pion) and 7 (proton), and the horizontal band. }}
    \label{fig:amp_and_effmass_stab}
\end{figure} 
 
In total we gather $\sim3000$ measurements of point-smeared (PS) and smeared-smeared (SS) correlators for each hadron and ensemble. Their analysis is performed using the Bayesian analysis framework for constrained curve fitting as described in \cite{Lepage:2002curve,Miller:2021omega}.  
The $n$-state function used as ansatz reads:
\begin{equation}
    C(t, Z_{P,n}, Z_{S,n}, E_n) = \sum_{n=0}^{n=N_s} Z_{S,n}Z_{S/P,n} e^{-E_nt}~~,
\end{equation}
where $Z_{P,n}$ and $Z_{S,n}$ are the amplitudes for the point and smeared sources/sinks. We are postponing a detailed study of the decay constants to future work and, therefore, do not include renormalisation constants at this point.
The fit is performed in a combined way on both the PS and SS correlators. Following \cite{Lepage:2002curve} the Bayesian constraints on the fit are introduced through an addition to the $\chi^2$ function to be minimized. We have to add a term for each prior, that is for every energy and the amplitude that we fix. The prior $\chi^2$ term is then:
\begin{equation}
    \chi^2_{prior} = \sum_{n=0}^{N_s} \frac{(Z_{P,n} - \tilde{Z}_{P,n})^2}{\tilde{\sigma}^2_{Z_{P,n}}} + \sum_{n=0}^{N_s} \frac{(Z_{S,n} - \tilde{Z}_{S,n})^2}{\tilde{\sigma}^2_{Z_{S,n}}} + \sum_{n=0}^{N_s} \frac{(E_{n} - \tilde{E}_{n})^2}{\tilde{\sigma}^2_{E_{n}}}~~,
\end{equation}
the chosen prior values for $E_0,Z_{P,n}, Z_{S,n}$ are normally distributed, while the excited-state energy priors are set to be log-normal, to preserve their order. 
The excited-state energy splittings are set to $2m_\pi$ with a width allowing for fluctuations down to one pion mass within one standard deviation.

\begin{figure}[t!]
    \centering
    \includegraphics[width=\textwidth]{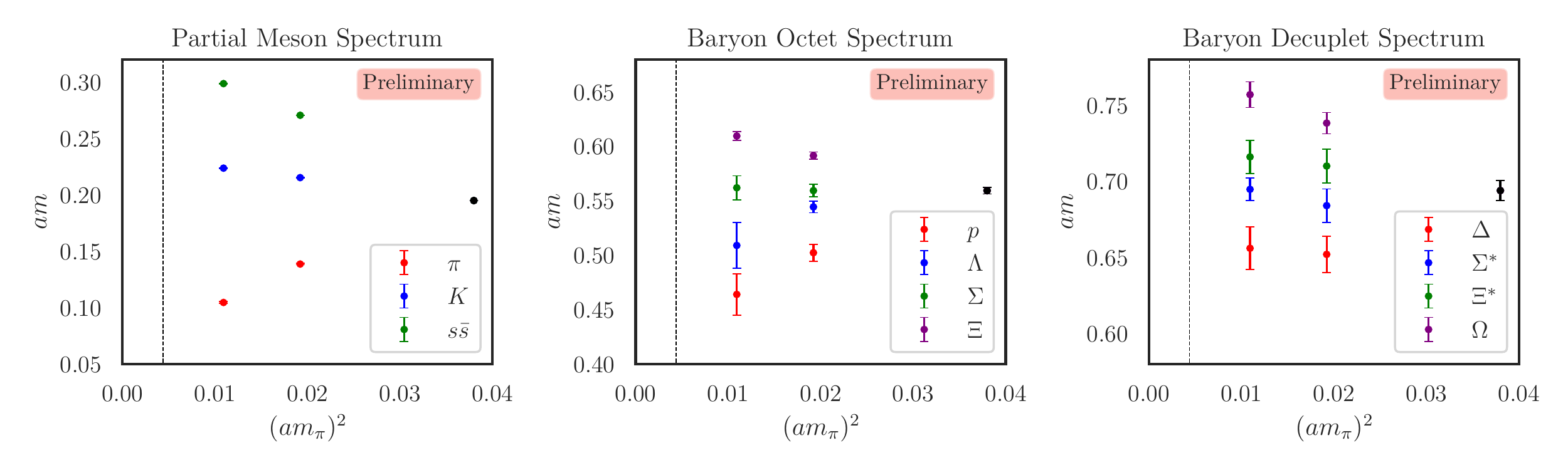}
    \caption{\textit{Spectrum for the light and strange hadrons for the ensembles at $\beta=3.8$. The left plot reports the mesons, the center plot is for the baryon octuplet, while the right one shows the baryon decuplet. The black point is the flavour symmetric ensemble a094m400. The vertical line represents the physical point.}}
    \label{fig:spectrum}
\end{figure}

In Fig.~\ref{fig:amp_and_effmass_stab} (top row) we present an overview of the prior setting procedure and the resulting final numbers for the proton on a094m300.
The choice of the standard deviation priors for the ground state amplitudes and energy has been set as roughly ten times the expected uncertainty of the final parameter, denoted by the light grey bands in the top row of the figure. 
To choose the fitting window ($t_{min}, t_{max}$) and the number of states to fit $N_s$ we perform a scan in the parameter space with fixed $t_{max}=28$ (pion) and 16 (proton).
The results for the pion (left) and proton (right) are shown in Fig.~\ref{fig:amp_and_effmass_stab} (bottom row). We observe stable fits with respect to the fit window and the number of states, provided $N_s$ is large enough to fully capture the correlation function. The choice of $t_{min}$ and $N_s$ is made on a per-ensemble basis for the mesons and the baryons separately, meaning that all mesons share the same set of parameters and all baryons share another.
In the future the new Bayesian model averaging \cite{Jay:2021modelavg} could be employed to further reduce this bias.

In passing we note that the $N_s=1,2$ results, corresponding to 1-state/2-state fits respectively, generally lead to comparable or smaller errors than the results obtained with the Bayesian method. 

The results for the masses of the light and strange hadrons obtained with the outlined procedure are summarised in Fig.~\ref{fig:spectrum}. In particular we show the meson, baryon octet and decuplet pion mass dependences. We are currently updating these studies to include all ensembles and configurations enabling a chiral and continuum extrapolation in the future. Our goal is to establish robust results as future benchmarks and standard reference for spectroscopy using SWF.

\section{The OPEN LATtice initiative}
\label{sec:openlat}

Stabilised Wilson fermions combine and build upon recent developments in the generation of gauge fields for use in lattice QCD. The newest ingredient, the exponentiated clover term, builds on long established paradigms of action design and one of its motivations was to make WCF safe for larger volumes. This includes in particular master-field type simulations, however their beneficial effect is there for all types and sizes of lattice.
The initial studies of \cite{Francis:2019muy} show a reduced $c_{SW}$ compared to standard clover simulations, alongside some indication of positive scaling behaviours. For example we saw a benefit in relative $am_q$ effects. In this current study we also observe good scaling in flow and hadronic observables going towards the continuum.
Deploying the full SWF toolkit we observe that stable simulations can be run in an extended parameter window.
The, unrelated, reported master-field simulations \cite{Fritzsch:2021klm,Ce:2021akh} are examples for their safety in $L$ but also the presented studies at $a=0.12$~fm in this work show the possible benefit. Recall, that large volume simulations with WCF are necessarily pathological as the clover term becomes non-invertible. 
How far SWF do extend the parameter window and what exactly the scaling benefits are is currently not determined. It is a key motivation for us to study SWF in more depth and to continue their investigation. 
To make this possible we founded and are announcing here a new collaborative effort: the OPEN LATtice initiative. It brings together researchers from different institutes and to pool resources. 
Together we want to generate state-of-the-art QCD gauge ensembles for physics applications and share them with the community under the open science philosophy.
Much of the progress achieved and reported here is under the umbrella of this initiative.

\subsection{Open science policy}

To us the goals of the initiative are closely tied to an open science policy. It is important that the community can access our results and gauge fields so that research on SWF can be facilitated and accelerated. 
We summarise our activity and open science policy in four main points:
\begin{itemize}
\itemsep0em
\item {\it define and uphold quality:} Define standards for control observables, continue to research and improve best practices. 
\item  {\it share and maintain repository:} Manage downloads and maintain data integrity, while making all control measurements and data available.
\item  {\it community boosting:} Use resource injections from members and interested/early access parties to expand set of gauges.
\item {\it  grant and enable access:} Ensure that configurations will be made open access with accompanying first publication without further discriminating access control. 		
\end{itemize}

\begin{figure}[t!]
\centering
\includegraphics[width=0.42\textwidth]{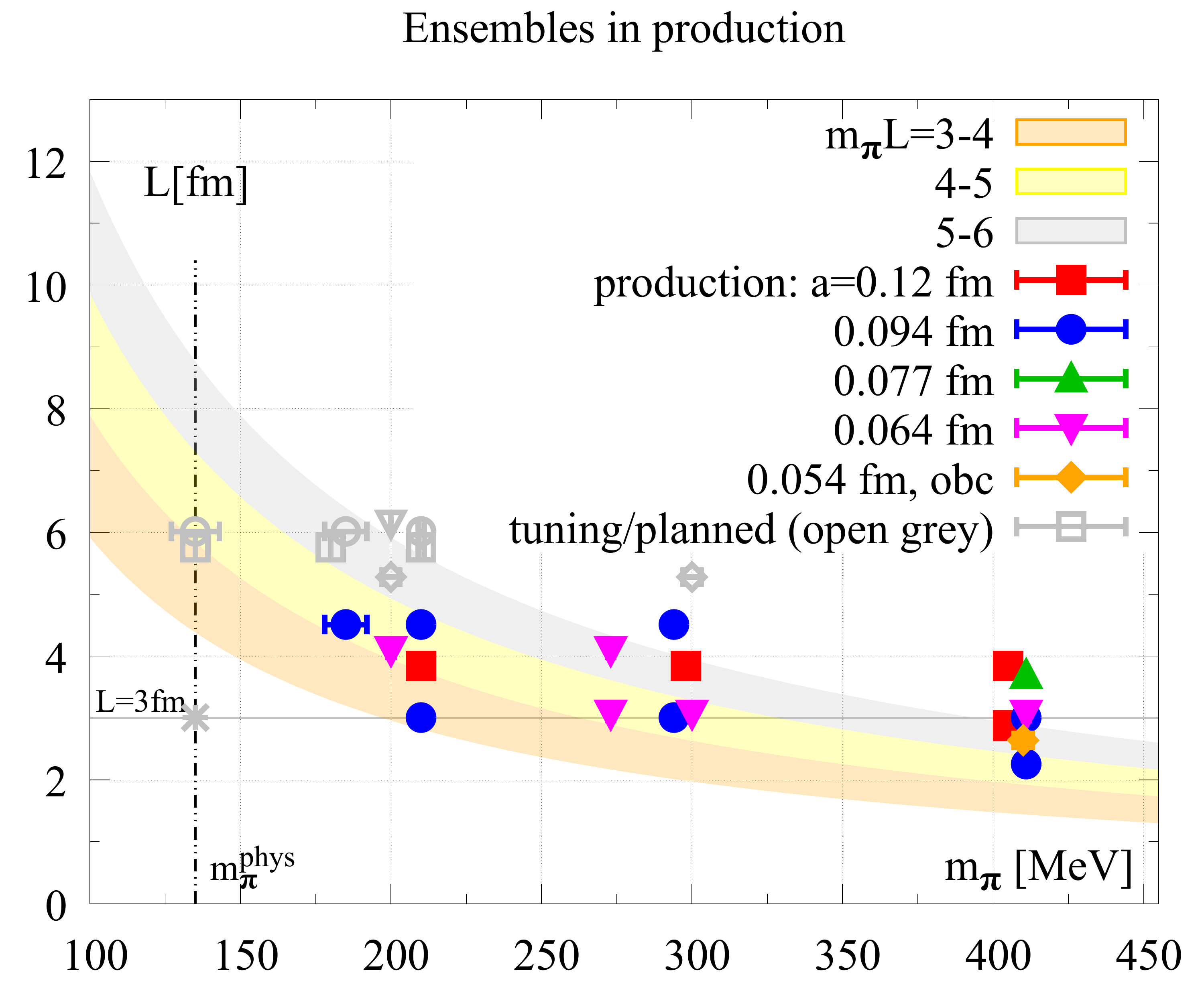}
\includegraphics[width=0.42\textwidth]{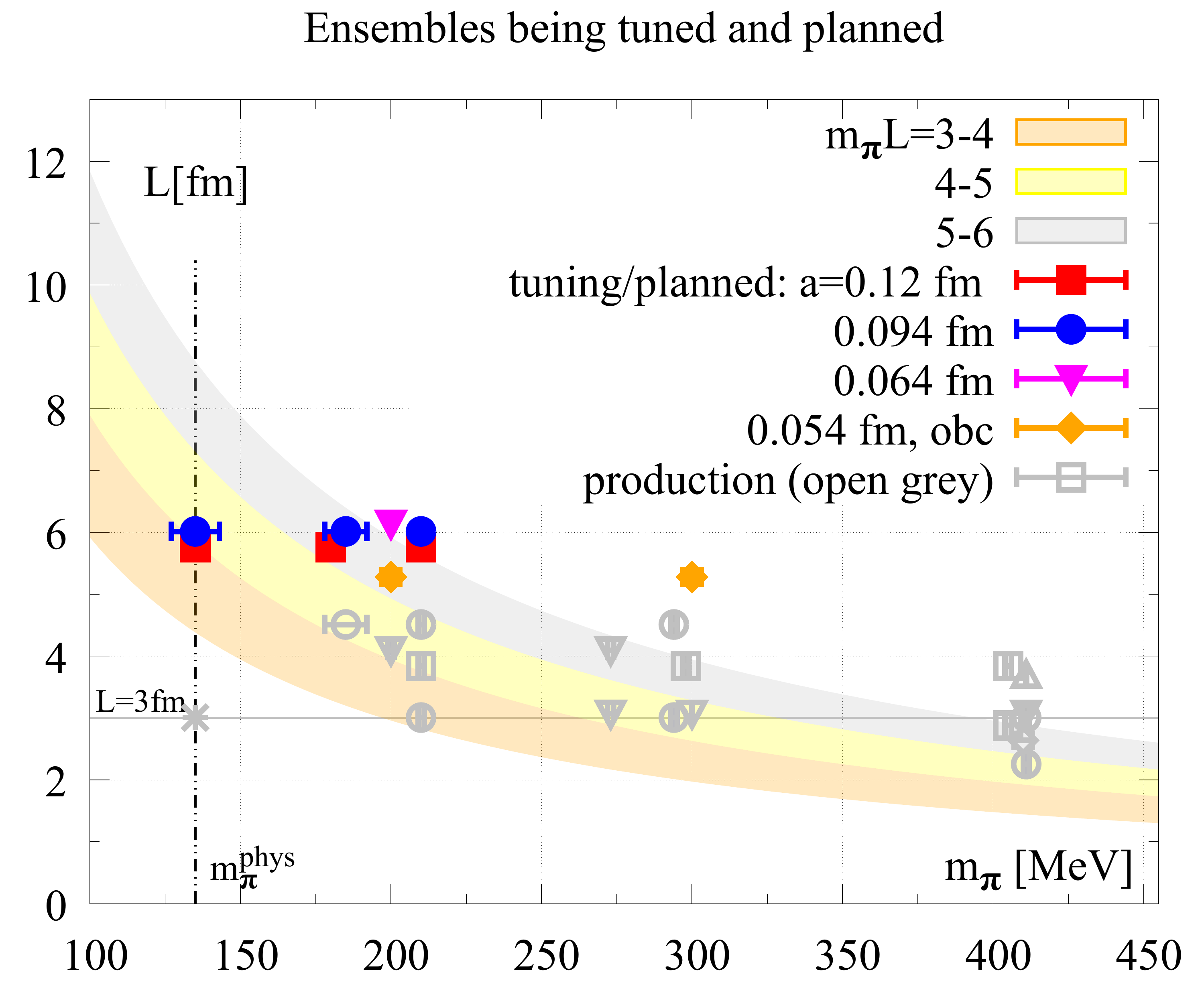}
\includegraphics[width=0.42\textwidth]{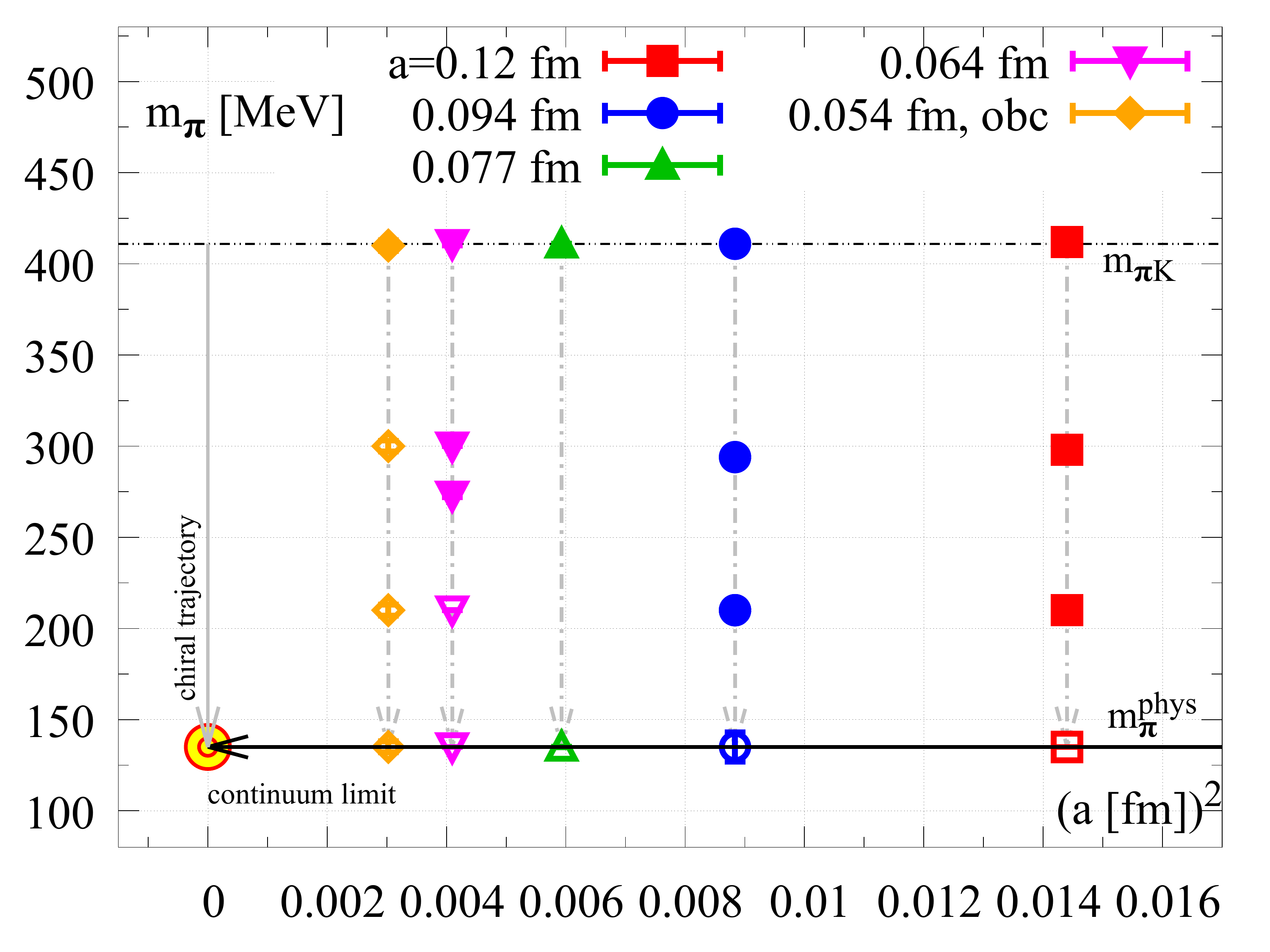}
\caption{\textit{Ensembles at the production (top-left) and at the tuning (top-right) stages. The colored bands denote the corresponding $m_\pi L$ regions. A discontinued a094 ensemble with $m_{\pi}L=2.1$ at $m_\pi=135$~MeV is shown as grey star. The bottom figure shows an overview of all ensembles over lattice spacing.}}
\label{fig:plans}
\end{figure}

\subsection{Planned ensembles}

Our goal is to generate ensembles that make best use of the beneficial SWF properties observed so far. First, we want to exploit the benefits in the coarse regime, where lighter pion masses than before seem possible. This is especially interesting for nucleon and nuclear physics applications where the cost of contractions can be high or signal-to-noise issues play a dominant role. 
Second, the ensembles should enable extrapolations from an as broad as possible window in $a$ and $m_\pi$ to control their systematics. Reduced mass dependent cutoff effects could prove useful, e.g., for applications in the meson sector like when studying $(g-2)_\mu$ or for the calculation of the the neutron electric dipole moment~\cite{Dragos:2019oxn}. Finally, they should cater for controlled estimation and exploitation of finite volume effects and their scaling. 

For all planned ensembles all configurations have to pass the criteria outlined in Sec.~\ref{sec:valid}. Additionally, they must satisfy the condition that $m_\pi L\gtrsim 4$ and $L\gtrsim 3$~fm for production level ensembles. Ideally all ensembles should be available with two volumes in the $m_\pi L \in 4 - 6$ range.

In the first iteration we plan to generate and share ensembles with four different lattice spacings, $a=0.064, 0.077, 0.094$ and $0.12$~fm, with periodic/anti-periodic boundary conditions. Furthermore, there will be one finer lattice spacing with open boundary conditions at $a=0.055$~fm\footnote{We use open boundary conditions as we observe signs of the onset of topology freezing at this lattice spacing.}.
Their simplified labels are a12, a094, a077, a064 and a055. The flavor symmetric point is tuned in all of them. 
Keeping in mind the flavor symmetric point is at $\sim412$~MeV the mass label is simplified to m400. The statistics goal for all ensembles is 500 independent configurations in the first, and 1000 independent configurations in the second, iteration.

With these ensembles at hand, the next step is to reduce the pion masses. Hinged on our previous studies with a094, our aim is to enable continuum limits at the same values of $m_\pi$. This entails tuning masses of $m_\pi \simeq 300$~MeV (label m300) and $m_\pi \simeq 210$~MeV (label m200). Once more, our statistics goal is 1000 independent configurations.
Finally, we aim for physical mass pions. With Wilson fermions this is a difficult goal, in particular at coarse lattice spacing, see e.g. \cite{Mohler:2020txx}. The SWF toolkit extends the parameter window, and whether it extends it this far is subject of our research. Our first jump towards the physical pion mass is being undertaken on the a094 line, where we have established the input parameters. We are currently studying the most stable algorithm, with a few configurations generated on the target spatial volume of $L=64$.

Regarding our ensembles it is important to distinguish clearly between the tuning and the production stage. In the tuning stage thermalisation is performed and the algorithm is adjusted. 
Once a stable setup is found, it is run for 100 independent configurations. If at this point the algorithm performs as expected and all quality criteria continue to be fulfilled, the run is declared production level. Then this first set of 100 configurations is also counted towards the total number of configurations. 
An overview plot of the planned ensembles is given in Fig.~\ref{fig:plans} (top). To make clear the distinction between production and tuning level we show the former on the left and the latter on the right. In the figure the pion mass is given as the $x$-axis while the physical volume is given as the $y$-axis. The lattice spacings are distinguished by the colors. Additionally, colored bands depicting the regimes $m_\pi L=3-4$, $4-5$ and $5-6$ are given for reference. In Fig.~\ref{fig:plans} (bottom) we give a slightly different visualisation of our planned ensembles now focusing on the continuum and chiral limits with the volume information suppressed. 


Any interested researcher is welcome to contact us to discuss how they can contribute. We want to thank those of you who have been in touch with us already and hope we can start new joint activities soon.

\section*{Acknowledgements}
The authors acknowledge support from the HPC computing centres and resources hpc-qcd (CERN),
HPE Apollo Hawk (HLRS) under the grant number stabwf/44185, Cori (NERSC), Frontera (TACC) as well as Occigen (CINES), Jean-Zay (IDRIS) and Ir\`ene-Joliot-Curie (TGCC) under projects 2020-A0080511504 and 2020-A0080502271 by GENCI. AS acknowledges funding support under the National Science Foundation grant PHY-1913287. The quenched results were obtained during our initial studies together with Martin L\"uscher, whose leading role we greatly acknowledge during the development and implementation of the SWF framework. 

\begin{spacing}{0.82}
\bibliographystyle{JHEP}
\bibliography{references}

\providecommand{\href}[2]{#2}\begingroup\raggedright\begin{thebibliography}{10}

\bibitem{Sheikholeslami:1985ij}
B.~Sheikholeslami and R.~Wohlert, \emph{{Improved Continuum Limit Lattice
  Action for QCD with Wilson Fermions}},
  \href{https://doi.org/10.1016/0550-3213(85)90002-1}{\emph{Nucl. Phys. B}
  {\bfseries 259} (1985) 572}.

\bibitem{Aoki:2021kgd}
Y.~Aoki et~al., \emph{{FLAG Review 2021}},
  \href{https://arxiv.org/abs/2111.09849}{{\ttfamily 2111.09849}}.

\bibitem{Frezzotti:2000nk}
{\scshape Alpha} collaboration, \emph{{Lattice QCD with a chirally twisted mass
  term}}, \href{https://doi.org/10.1088/1126-6708/2001/08/058}{\emph{JHEP}
  {\bfseries 08} (2001) 058}
  [\href{https://arxiv.org/abs/hep-lat/0101001}{{\ttfamily hep-lat/0101001}}].

\bibitem{Frezzotti:2005gi}
R.~Frezzotti, G.~Martinelli, M.~Papinutto and G.C.~Rossi, \emph{{Reducing
  cutoff effects in maximally twisted lattice QCD close to the chiral limit}},
  \href{https://doi.org/10.1088/1126-6708/2006/04/038}{\emph{JHEP} {\bfseries
  04} (2006) 038} [\href{https://arxiv.org/abs/hep-lat/0503034}{{\ttfamily
  hep-lat/0503034}}].

\bibitem{Shindler:2007vp}
A.~Shindler, \emph{{Twisted mass lattice QCD}},
  \href{https://doi.org/10.1016/j.physrep.2008.03.001}{\emph{Phys. Rept.}
  {\bfseries 461} (2008) 37} [\href{https://arxiv.org/abs/0707.4093}{{\ttfamily
  0707.4093}}].

\bibitem{Luscher:2017cjh}
M.~L\"uscher, \emph{{Stochastic locality and master-field simulations of very
  large lattices}},
  \href{https://doi.org/10.1051/epjconf/201817501002}{\emph{EPJ Web Conf.}
  {\bfseries 175} (2018) 01002}
  [\href{https://arxiv.org/abs/1707.09758}{{\ttfamily 1707.09758}}].

\bibitem{Francis:2019muy}
A.~Francis, P.~Fritzsch, M.~L\"uscher and A.~Rago, \emph{{Master-field
  simulations of O($a$)-improved lattice QCD: Algorithms, stability and
  exactness}}, \href{https://doi.org/10.1016/j.cpc.2020.107355}{\emph{Comput.
  Phys. Commun.} {\bfseries 255} (2020) 107355}
  [\href{https://arxiv.org/abs/1911.04533}{{\ttfamily 1911.04533}}].

\bibitem{Ce:2021akh}
M.~C\`e, M.~Bruno, J.~Bulava, A.~Francis, P.~Fritzsch, J.R.~Green et~al.,
  \emph{{Approaching the master-field: Hadronic observables in large volumes}},
   10, 2021 [\href{https://arxiv.org/abs/2110.15375}{{\ttfamily 2110.15375}}].

\bibitem{Fritzsch:2021klm}
P.~Fritzsch, J.~Bulava, M.~C\`e, A.~Francis, M.~L\"uscher and A.~Rago,
  \emph{{Master-field simulations of QCD}},  11, 2021
  [\href{https://arxiv.org/abs/2111.11544}{{\ttfamily 2111.11544}}].

\bibitem{mluscher:openqcd}
M.~L\"uscher, \emph{{Code available at
  \url{https://luscher.web.cern.ch/luscher/openQCD/}}}, .

\bibitem{Horowitz:1985kd}
A.M.~Horowitz, \emph{{Stochastic Quantization in Phase Space}},
  \href{https://doi.org/10.1016/0370-2693(85)91360-7}{\emph{Phys. Lett. B}
  {\bfseries 156} (1985) 89}.

\bibitem{Horowitz:1986dt}
A.M.~Horowitz, \emph{{The Second Order Langevin Equation and Numerical
  Simulations}},
  \href{https://doi.org/10.1016/0550-3213(87)90159-3}{\emph{Nucl. Phys. B}
  {\bfseries 280} (1987) 510}.

\bibitem{HOROWITZ1991247}
A.M.~Horowitz, \emph{A generalized guided monte carlo algorithm},
  \href{https://doi.org/https://doi.org/10.1016/0370-2693(91)90812-5}{\emph{Physics
  Letters B} {\bfseries 268} (1991) 247}.

\bibitem{Jansen:1995gz}
K.~Jansen and C.~Liu, \emph{{Kramers equation algorithm for simulations of QCD
  with two flavors of Wilson fermions and gauge group SU(2)}},
  \href{https://doi.org/10.1016/0550-3213(95)00427-T}{\emph{Nucl. Phys. B}
  {\bfseries 453} (1995) 375}
  [\href{https://arxiv.org/abs/hep-lat/9506020}{{\ttfamily hep-lat/9506020}}].

\bibitem{mluscher:note}
M.~L\"uscher, \emph{{Ergodicity of the SMD algorithm in lattice QCD,
  unpublished notes (2017)},
  \url{http://luscher.web.cern.ch/luscher/notes/smd-ergodicity.pdf}}, .

\bibitem{Luscher:2011kk}
M.~L\"uscher and S.~Schaefer, \emph{{Lattice QCD without topology barriers}},
  \href{https://doi.org/10.1007/JHEP07(2011)036}{\emph{JHEP} {\bfseries 07}
  (2011) 036} [\href{https://arxiv.org/abs/1105.4749}{{\ttfamily 1105.4749}}].

\bibitem{Orginos:2017kos}
K.~Orginos, A.~Radyushkin, J.~Karpie and S.~Zafeiropoulos, \emph{{Lattice QCD
  exploration of parton pseudo-distribution functions}},
  \href{https://doi.org/10.1103/PhysRevD.96.094503}{\emph{Phys. Rev. D}
  {\bfseries 96} (2017) 094503}
  [\href{https://arxiv.org/abs/1706.05373}{{\ttfamily 1706.05373}}].

\bibitem{Shanahan:2020zxr}
P.~Shanahan, M.~Wagman and Y.~Zhao, \emph{{Collins-Soper kernel for TMD
  evolution from lattice QCD}},
  \href{https://doi.org/10.1103/PhysRevD.102.014511}{\emph{Phys. Rev. D}
  {\bfseries 102} (2020) 014511}
  [\href{https://arxiv.org/abs/2003.06063}{{\ttfamily 2003.06063}}].

\bibitem{Shindler:2015aqa}
A.~Shindler, T.~Luu and J.~de~Vries, \emph{{Nucleon electric dipole moment with
  the gradient flow: The \ensuremath{\theta}-term contribution}},
  \href{https://doi.org/10.1103/PhysRevD.92.094518}{\emph{Phys. Rev. D}
  {\bfseries 92} (2015) 094518}
  [\href{https://arxiv.org/abs/1507.02343}{{\ttfamily 1507.02343}}].

\bibitem{Altenkort:2020fgs}
L.~Altenkort, A.M.~Eller, O.~Kaczmarek, L.~Mazur, G.D.~Moore and H.-T.~Shu,
  \emph{{Heavy quark momentum diffusion from the lattice using gradient flow}},
  \href{https://doi.org/10.1103/PhysRevD.103.014511}{\emph{Phys. Rev. D}
  {\bfseries 103} (2021) 014511}
  [\href{https://arxiv.org/abs/2009.13553}{{\ttfamily 2009.13553}}].

\bibitem{Lucini:2021xke}
B.~Lucini, E.~Bennett, J.~Holligan, D.K.~Hong, H.~Hsiao, J.-W.~Lee et~al.,
  \emph{{Sp(4) gauge theories and beyond the standard modelphysics}},  11, 2021
  [\href{https://arxiv.org/abs/2111.12125}{{\ttfamily 2111.12125}}].

\bibitem{Brambilla:2020siz}
N.~Brambilla, V.~Leino, P.~Petreczky and A.~Vairo, \emph{{Lattice QCD
  constraints on the heavy quark diffusion coefficient}},
  \href{https://doi.org/10.1103/PhysRevD.102.074503}{\emph{Phys. Rev. D}
  {\bfseries 102} (2020) 074503}
  [\href{https://arxiv.org/abs/2007.10078}{{\ttfamily 2007.10078}}].

\bibitem{Altenkort:2020axj}
L.~Altenkort, A.M.~Eller, O.~Kaczmarek, L.~Mazur, G.D.~Moore and H.-T.~Shu,
  \emph{{Sphaleron rate from Euclidean lattice correlators: An exploration}},
  \href{https://doi.org/10.1103/PhysRevD.103.114513}{\emph{Phys. Rev. D}
  {\bfseries 103} (2021) 114513}
  [\href{https://arxiv.org/abs/2012.08279}{{\ttfamily 2012.08279}}].

\bibitem{Luscher:1996ug}
M.~L\"uscher, S.~Sint, R.~Sommer, P.~Weisz and U.~Wolff, \emph{{Nonperturbative
  O(a) improvement of lattice QCD}},
  \href{https://doi.org/10.1016/S0550-3213(97)00080-1}{\emph{Nucl. Phys. B}
  {\bfseries 491} (1997) 323}
  [\href{https://arxiv.org/abs/hep-lat/9609035}{{\ttfamily hep-lat/9609035}}].

\bibitem{Giusti:2018cmp}
L.~Giusti and M.~L\"uscher, \emph{{Topological susceptibility at $T>T_{\rm c}$
  from master-field simulations of the SU(3) gauge theory}},
  \href{https://doi.org/10.1140/epjc/s10052-019-6706-7}{\emph{Eur. Phys. J. C}
  {\bfseries 79} (2019) 207}
  [\href{https://arxiv.org/abs/1812.02062}{{\ttfamily 1812.02062}}].

\bibitem{Bruno:2016plf}
M.~Bruno, T.~Korzec and S.~Schaefer, \emph{{Setting the scale for the CLS $2 +
  1$ flavor ensembles}},
  \href{https://doi.org/10.1103/PhysRevD.95.074504}{\emph{Phys. Rev. D}
  {\bfseries 95} (2017) 074504}
  [\href{https://arxiv.org/abs/1608.08900}{{\ttfamily 1608.08900}}].

\bibitem{Bulava:2013cta}
J.~Bulava and S.~Schaefer, \emph{{Improvement of $N_f$ = 3 lattice QCD with
  Wilson fermions and tree-level improved gauge action}},
  \href{https://doi.org/10.1016/j.nuclphysb.2013.05.019}{\emph{Nucl. Phys. B}
  {\bfseries 874} (2013) 188}
  [\href{https://arxiv.org/abs/1304.7093}{{\ttfamily 1304.7093}}].

\bibitem{Bietenholz:2010jr}
W.~Bietenholz et~al., \emph{{Tuning the strange quark mass in lattice
  simulations}},
  \href{https://doi.org/10.1016/j.physletb.2010.05.067}{\emph{Phys. Lett. B}
  {\bfseries 690} (2010) 436}
  [\href{https://arxiv.org/abs/1003.1114}{{\ttfamily 1003.1114}}].

\bibitem{Bruno:2014jqa}
M.~Bruno et~al., \emph{{Simulation of QCD with N$_{f} =$ 2 $+$ 1 flavors of
  non-perturbatively improved Wilson fermions}},
  \href{https://doi.org/10.1007/JHEP02(2015)043}{\emph{JHEP} {\bfseries 02}
  (2015) 043} [\href{https://arxiv.org/abs/1411.3982}{{\ttfamily 1411.3982}}].

\bibitem{Aoki:2016frl}
S.~Aoki et~al., \emph{{Review of lattice results concerning low-energy particle
  physics}}, \href{https://doi.org/10.1140/epjc/s10052-016-4509-7}{\emph{Eur.
  Phys. J. C} {\bfseries 77} (2017) 112}
  [\href{https://arxiv.org/abs/1607.00299}{{\ttfamily 1607.00299}}].

\bibitem{Luscher:2013cpa}
M.~L\"uscher, \emph{{Chiral symmetry and the Yang--Mills gradient flow}},
  \href{https://doi.org/10.1007/JHEP04(2013)123}{\emph{JHEP} {\bfseries 04}
  (2013) 123} [\href{https://arxiv.org/abs/1302.5246}{{\ttfamily 1302.5246}}].

\bibitem{Walker-Loud:2008rui}
A.~Walker-Loud et~al., \emph{{Light hadron spectroscopy using domain wall
  valence quarks on an Asqtad sea}},
  \href{https://doi.org/10.1103/PhysRevD.79.054502}{\emph{Phys. Rev. D}
  {\bfseries 79} (2009) 054502}
  [\href{https://arxiv.org/abs/0806.4549}{{\ttfamily 0806.4549}}].

\bibitem{Walker-Loud:2014iea}
A.~Walker-Loud, \emph{{Nuclear Physics Review}},
  \href{https://doi.org/10.22323/1.187.0013}{\emph{PoS} {\bfseries LATTICE2013}
  (2014) 013} [\href{https://arxiv.org/abs/1401.8259}{{\ttfamily 1401.8259}}].

\bibitem{NPLQCD:2010ocs}
{\scshape NPLQCD} collaboration, \emph{{Evidence for a Bound H-dibaryon from
  Lattice QCD}},
  \href{https://doi.org/10.1103/PhysRevLett.106.162001}{\emph{Phys. Rev. Lett.}
  {\bfseries 106} (2011) 162001}
  [\href{https://arxiv.org/abs/1012.3812}{{\ttfamily 1012.3812}}].

\bibitem{Inoue:2010es}
{\scshape HAL QCD} collaboration, \emph{{Bound H-dibaryon in Flavor SU(3) Limit
  of Lattice QCD}},
  \href{https://doi.org/10.1103/PhysRevLett.106.162002}{\emph{Phys. Rev. Lett.}
  {\bfseries 106} (2011) 162002}
  [\href{https://arxiv.org/abs/1012.5928}{{\ttfamily 1012.5928}}].

\bibitem{Francis:2018qch}
A.~Francis, J.R.~Green, P.M.~Junnarkar, C.~Miao, T.D.~Rae and H.~Wittig,
  \emph{{Lattice QCD study of the $H$ dibaryon using hexaquark and two-baryon
  interpolators}},
  \href{https://doi.org/10.1103/PhysRevD.99.074505}{\emph{Phys. Rev. D}
  {\bfseries 99} (2019) 074505}
  [\href{https://arxiv.org/abs/1805.03966}{{\ttfamily 1805.03966}}].

\bibitem{Green:2021qol}
J.R.~Green, A.D.~Hanlon, P.M.~Junnarkar and H.~Wittig, \emph{{Weakly bound $H$
  dibaryon from SU(3)-flavor-symmetric QCD}},
  \href{https://arxiv.org/abs/2103.01054}{{\ttfamily 2103.01054}}.

\bibitem{Green:2021sxb}
J.R.~Green, A.D.~Hanlon, P.M.~Junnarkar and H.~Wittig, \emph{{Continuum limit
  of baryon-baryon scattering with SU(3) flavor symmetry}},  in \emph{{38th
  International Symposium on Lattice Field Theory}}, 11, 2021
  [\href{https://arxiv.org/abs/2111.09675}{{\ttfamily 2111.09675}}].

\bibitem{callat:lalibe}
A.~Gambhir, D.~Brantley, J.~Chang, B.~Hörz, H.~Monge-Camacho, P.~Vranas
  et~al., \emph{{Code available at
  \url{https://github.com/callat-qcd/lalibe}}}, .

\bibitem{Edwards:2004sx}
{\scshape SciDAC, LHPC, UKQCD} collaboration, \emph{{The Chroma software system
  for lattice QCD}},
  \href{https://doi.org/10.1016/j.nuclphysbps.2004.11.254}{\emph{Nucl. Phys. B
  Proc. Suppl.} {\bfseries 140} (2005) 832}
  [\href{https://arxiv.org/abs/hep-lat/0409003}{{\ttfamily hep-lat/0409003}}].

\bibitem{Clark:2009wm}
M.A.~Clark, R.~Babich, K.~Barros, R.C.~Brower and C.~Rebbi, \emph{{Solving
  Lattice QCD systems of equations using mixed precision solvers on GPUs}},
  \href{https://doi.org/10.1016/j.cpc.2010.05.002}{\emph{Comput. Phys. Commun.}
  {\bfseries 181} (2010) 1517}
  [\href{https://arxiv.org/abs/0911.3191}{{\ttfamily 0911.3191}}].

\bibitem{Gusken:1990smear}
S.~Güsken, \emph{A study of smearing techniques for hadron correlation
  functions},
  \href{https://doi.org/https://doi.org/10.1016/0920-5632(90)90273-W}{\emph{Nuclear
  Physics B - Proceedings Supplements} {\bfseries 17} (1990) 361}.

\bibitem{Lepage:2002curve}
G.~Lepage, B.~Clark, C.~Davies, K.~Hornbostel, P.~Mackenzie, C.~Morningstar
  et~al., \emph{Constrained curve fitting},
  \href{https://doi.org/https://doi.org/10.1016/S0920-5632(01)01638-3}{\emph{Nuclear
  Physics B - Proceedings Supplements} {\bfseries 106-107} (2002) 12}.

\bibitem{Miller:2021omega}
N.~Miller, L.~Carpenter, E.~Berkowitz, C.C.~Chang, B.~H\"orz, D.~Howarth
  et~al., \emph{Scale setting the m\"obius domain wall fermion on
  gradient-flowed hisq action using the omega baryon mass and the gradient-flow
  scales ${t}_{0}$ and ${w}_{0}$},
  \href{https://doi.org/10.1103/PhysRevD.103.054511}{\emph{Phys. Rev. D}
  {\bfseries 103} (2021) 054511}.

\bibitem{Jay:2021modelavg}
W.I.~Jay and E.T.~Neil, \emph{Bayesian model averaging for analysis of lattice
  field theory results},
  \href{https://doi.org/10.1103/PhysRevD.103.114502}{\emph{Phys. Rev. D}
  {\bfseries 103} (2021) 114502}.

\bibitem{Dragos:2019oxn}
J.~Dragos, T.~Luu, A.~Shindler, J.~de~Vries and A.~Yousif, \emph{{Confirming
  the Existence of the strong CP Problem in Lattice QCD with the Gradient
  Flow}}, \href{https://doi.org/10.1103/PhysRevC.103.015202}{\emph{Phys. Rev.
  C} {\bfseries 103} (2021) 015202}
  [\href{https://arxiv.org/abs/1902.03254}{{\ttfamily 1902.03254}}].

\bibitem{Mohler:2020txx}
D.~Mohler and S.~Schaefer, \emph{{Remarks on strange-quark simulations with
  Wilson fermions}},
  \href{https://doi.org/10.1103/PhysRevD.102.074506}{\emph{Phys. Rev. D}
  {\bfseries 102} (2020) 074506}
  [\href{https://arxiv.org/abs/2003.13359}{{\ttfamily 2003.13359}}].

\end{thebibliography}\endgroup
\end{spacing}


\end{document}